\documentclass[prd,aps,twocolumn,floats,floatfix,nofootinbib]{revtex4-1}

\usepackage{amssymb,bm}
\usepackage{graphicx}
\usepackage{dcolumn}
\usepackage{bm}
\usepackage{graphics}
\usepackage{slashed}
\usepackage{enumerate}
\usepackage{amsmath}
\usepackage{url}
\usepackage[usenames,dvipsnames,svgnames,table]{xcolor}
\usepackage{color}
\usepackage{xcolor}

\definecolor{color1}{RGB}{191, 0, 255}

\begin{document}

\title{Turbulent magnetic-field amplification in the first 10 milliseconds after a binary neutron star merger: comparing high-resolution and large eddy simulations}

\author{
Ricard Aguilera-Miret$^{1,2,3}$,
Daniele Vigan\`o$^{2,3,4}$,
Federico Carrasco$^{2,5}$,
Borja Mi\~nano$^{1,2,3}$,
Carlos Palenzuela$^{1,2,3}$
}

\affiliation{${^1}$Departament  de  F\'{\i}sica $\&$ IAC3,  Universitat  de  les  Illes  Balears  and  Institut  d'Estudis
Espacials  de  Catalunya,  Palma  de  Mallorca,  Baleares  E-07122,  Spain\\
$^2$Institut Aplicacions Computationals (IAC3),  Universitat  de  les  Illes  Balears,  Palma  de  Mallorca,  Baleares  E-07122,  Spain\\
$^3$Institut d'Estudis Espacials de Catalunya (IEEC), 08034 Barcelona, Spain\\
$^4$Institute of Space Sciences (ICE, CSIC), 08193 Barcelona, Spain\\
$^5$Max Planck Institute for Gravitational Physics, 14476 Potsdam, Germany}
 
\begin{abstract}
	
The detection of binary neutron star mergers represents one of the most important and complex astrophysical discoveries of the recent years. One of the unclear aspects of the problem is the turbulent magnetic field amplification, initially triggered by the Kelvin-Helmholtz instability at much smaller scales than any reachable numerical resolution nowadays. Here we present numerical simulations of the first ten milliseconds of a binary neutron star merger. First, we confirm in detail how the simulated amplification depends on the numerical resolution and is distributed on a broad range of scales, as expected from turbulent MHD theory. 
We find that an initial large-scale magnetic field of $10^{11}\,$G inside each star is amplified in the remnant to root-mean-square values above $10^{16}\,$G within the first $5$ milliseconds for our highest-resolution run.
Then, we run large eddy simulations, exploring the performance of the subgrid-scale gradient model, already tested successfully in previous turbulent box simulations. 
We show that the addition of this model is especially important in the induction equation, since it leads to an amplification of the magnetic field comparable to a higher-resolution run, but with a greatly reduced computational cost. In the first 10 milliseconds, there is no clear hint for an ordered, large-scale magnetic field, which should indeed occur in longer timescales through magnetic winding and the magneto-rotational instability.
\end{abstract}

\maketitle

\section{Introduction}

The extraordinary multimessenger observations of GW170817 \cite{abbott17a,abbott17b} demonstrated that binary neutron star (BNS) mergers can produce strong gravitational waves (GW) signals and power bright electromagnetic (EM) emissions across the spectrum \cite{goldstein2017,savchenko2017,abbott17d,abbott17c,metzger17,davanzo2018,fong2019,dobie2018,mooley2018}.
These signals have already served to put some constraints on the physical properties of neutron stars (NSs) (see, e.g., \cite{margalit17,shibata2017modeling,abbott2018}), such as their radius and maximum mass, tidal deformability and equation-of-state (EoS), among others.

Although the central aspects of BNS systems are qualitatively understood, the details of the merger and post-merger dynamics remain only poorly constrained, with many important questions still open. In this paper we are mainly concerned with one of such issues: the amplification and large-scale (re-)organization of the magnetic field, arguably required to launch the successful jet outflows associated to the short gamma-ray burst (SGRB).
Despite the recent progress of general-relativistic magnetohydrodynamics (GRMHD) simulations \cite{palenzuela2013electromagnetic,kiuchi14,neilsen2014magnetized,kiuchi15,giacomazzo15,palenzuela15,ruiz16,kiuchi18,ciolfi2019,ciolfi2020collimated,ruiz2020,mosta2020}, the impact of magnetic turbulence on the evolution of the hypermassive neutron star (HMNS) remnant is highly uncertain, mostly due to the lack of a spatial resolution able to capture all the relevant scales. 
It has been recognized that the effects of turbulent viscosity and dynamo (so far numerically under-resolved), along with neutrino transport, can be crucial for the redistribution of angular momentum, mass ejecta, lifetime of the remnant and production of the jet (e.g., \cite{ciolfi2020key}). 

Observationally, the typical range of magnetic field strengths characterizing Gyr-old NSs (typical age at which binaries can merge) is $10^{8-11}\,$G \cite{tauris17}.\footnote{These values refer to the dipolar component at the NS surface, while 
stronger field (by one order of magnitude or more) could be expected at their interiors or due to higher multipole components (see, e.g., \cite{bilous2019nicer,rea2010low}).}   
Magnetic field amplification occurs during and after merger through a number of distinct MHD mechanisms, channeling a fraction of the abundant orbital kinetic energy ($\sim10^{53}\,$erg) of the system.
The Kelvin-Helmholtz instability (KHI), originated in the shearing layer at the collision interface, drastically enhances the magnetic field by stretching and folding embedded field lines in a process known as small-scale turbulent dynamo. 
Local special-relativistic MHD simulations have shown that the development of the KHI at merger can generate magnetar-level magnetic field strengths within the first few milliseconds \cite{obergaulinger10,zrake13b}.
Later, GRMHD simulations of BNS mergers of unprecedented high-resolution (grid-spacing of $17.5\,$m) \cite{kiuchi15} showed that an initial magnetic field of moderate strength $10^{13}\,$G can be amplified up to $\sim10^{16}\,$G within $\sim5\,$ms after merger, reaching magnetic saturation levels at energies $E_B \gtrsim 10^{50}\,$erg. However, no sign of numerical convergence was found, meaning that the KHI is not yet fully resolved even at those resolutions.

After the quick growth of the magnetic field due to the KHI, there are two other mechanisms associated to the differentially rotating HMNS that dominates on longer timescales $\gtrsim10\,$ms: magnetic winding, which linearly amplifies the toroidal components of the field from the poloidal ones, and the magneto-rotational instability (MRI). For the latter, the wavelengths of the fastest growing modes are proportional to the magnetic fields. Therefore, even the highest-resolution GRMHD simulations to date cannot resolve the MRI, 
unless artificially large initial magnetic fields above $10^{13}\,$G are adopted as to increase the associated cutoff length scales. Even in this way, simulations are
far from capturing the turbulent cascade all the way down to the viscous scale (determined by neutrino viscosity \cite{guilet17}), as it would be required for a direct numerical simulation (DNS). Finally, efficient MRI amplification is expected to continue acting inside the accretion disk after the remnant collapses to a black hole.

In the absence of computationally viable DNS to consistently evolve all the phases of the magnetic dynamics described above, different approaches were considered. Many studies  have imposed rather large initial pre-merger (e.g.,\cite{ruiz16,kiuchi18,ciolfi2019,ciolfi2020collimated,ruiz2020}) or post-merger (\cite{mosta2020}) magnetic field strengths $\sim10^{14-16}\,$G, to compensate the inability to capture the KHI amplification. However, the quantitative results may not be fully reliable, since the amplification via KHI happens over a broad range of scales and does not preserve a large-scale ordered field.

One of the most promising alternatives is performing large eddy simulations (LESs), in which the evolution equations are modified in order to account for the unresolved subgrid-scale (SGS) dynamics \cite{zhiyin15}. This method was applied, in the present context, by including new terms (chosen proportional to the fluid vorticity) into the induction equation \cite{bucciantini13,giacomazzo15,palenzuela15}. While the results of these studies show an effective growth of the magnetic field, they do not match the physical MHD dynamics and rely on arbitrarily tuning and switching ``by hand'' of the extra terms. Other approaches have, instead, centered their attention on the turbulent viscous effect during the post-merger phase, evolving viscous hydrodynamics (HD) in substitution of the MHD equations \cite{duez2004,shibata2017general,radice17,fujibayashi2020}. These models are however unable, by construction, to capture the dynamo mechanism and depend on parameters to be calibrated via very high resolution GRMHD simulations (e.g., \cite{radice2020}).

A more sophisticated alternative, based on the so-called gradient SGS model \cite{leonard75,muller02a}, has been proposed recently for Newtonian, special and general relativistic MHD, respectively in \cite{vigano19b,carrasco19,vigano20}. It was proven to have very good performance (in terms of capturing the magnetic amplification especially) in box simulations of the KHI, for a variety of initial conditions and resolutions, but it was not yet implemented in BNS mergers. The advantage of this approach is that it relies on the mathematical expansion of the fields involved in the dynamics, with no a-priori physical assumptions. In that sence, this SGS model is conceptually similar to high-order reconstruction methods used in finite-volume numerical schemes. 

In this paper we perform BNS merger simulations, focusing on the magnetic field amplification due to the KHI during the first $\sim10\,$ms after merger.
In contrast to previous studies, our simulations begin with each star having realistic magnetic field strength values of about $10^{11}\,$G.  We use high-order numerical methods and the elaborated gradient SGS model already presented for GRMHD box simulations of the KHI \cite{vigano20}, which is applied for the first time to the BNS merger scenario.

This article is organized as follows: our LES approach for GRMHD is briefly revisited on \S\ref{sec:equations}. The general setup, as well as the numerical methods, is described on \S\ref{sec:setup}. The results of the simulations are presented and analyzed in \S\ref{sec:results}. Conclusions are drawn on \S\ref{sec:conclusions}.

\section{Large eddy simulations in GRMHD}\label{sec:equations}

The concept and the mathematical foundations behind the explicit LES with a gradient SGS approach have been extensively explained in our previous works (and references within) in the context of Newtonian~\cite{vigano19b} and relativistic MHD \cite{carrasco19,vigano20}, to which we refer for details and further previous references. In brief, the space discretization in any numerical simulation can be seen as a filtering of the continuous solution, with an implicit kernel (numerical-method-dependent) having the size of the numerical grid, $\Delta$. The evolved numerical values of the fields can be then formally interpreted as weighted averages (or filtered) over the numerical cell. Seen in this way, the subgrid deviations of the field values from their averages causes a loss of information at small scales, for those terms which are nonlinear functions of the evolved variables. SGS terms obtained from the gradient model are added to the equations in order to partially compensate such loss.

Under the $3+1$ decomposition framework~\cite{bonabook}, the line element can be written as
\begin{equation}
ds^2 = - \alpha^2 \, dt^2 + \gamma_{ij} \bigl( dx^i + \beta^i dt \bigr) \bigl( dx^j + \beta^j dt \bigr)~, 
\label{3+1decom}  
\end{equation}
where $\alpha$ is the lapse function, $\beta^{i}$ is the shift vector, and $\gamma_{ij}$ is the induced metric on each spatial foliation, with determinant $\sqrt{\gamma}$. We use the covariant conformal Z4 formulation~\cite{alic12,bezares17} to evolve the Einstein equations. A summary of the final set of evolution equation for the spacetime fields, together with the gauge conditions setting the choice of coordinates, can be found e.g. in~\cite{palenzuela18}.

The GRMHD equations for a magnetized, non-viscous and perfectly conducting fluid \cite{palenzuela15} (in units $G=c=M_{\odot}=1$) consider the set of conserved variables $ \left\lbrace \sqrt{\gamma}D, \sqrt{\gamma}S^i , \sqrt{\gamma}U, \sqrt{\gamma}B^i \right\rbrace$. They are functions of the rest-mass density $\rho$, the specific internal energy $\epsilon$, the velocity vector $v^{i}$ and the magnetic field $B^i$ (primitive fields), as follows:
\begin{eqnarray}
D &=& \rho W ~, \\
S^i &=& (h W^2 + B^2 ) v^i - (B^k v_k) B^i ~,\\
U &=& h W^2 - p + B^2 - \frac{1}{2} \left[ (B^k v_k)^2 + \frac{B^2}{W^2} \right]  ~,
\end{eqnarray}
where $W = (1-v^2)^{-1/2}$ is the Lorentz factor. The pressure $p$ is defined through the EoS detailed in \S\ref{sec:setup}. The discretized evolution equations, including the hyperbolic divergence cleaning via damping of the field $\phi$ \citep{palenzuela18}, can be written as follows:\footnote{Comparing with our previous works~\cite{vigano19b,carrasco19,vigano20} where the entire formalism was presented, we have hereafter simplified the notation by removing the tildes and bars from the filtered fields and fluxes, for the sake of clarity. All fields in the equations are implicitly meant to be the filtered values (i.e., simply resolved by the discretized equations, as in any simulation).}

\begin{eqnarray}
&& \partial_t (\sqrt{\gamma} D) + \partial_k [- \beta^k \sqrt{\gamma} D + \alpha \sqrt{\gamma} ({N}^k - {\tau}^{k}_{N})] = 0 ~,
\label{evol_D_sgs} \nonumber \\
&& \partial_t (\sqrt{\gamma} {S}_i) + \partial_k [- \beta^k \sqrt{\gamma} {{S}}_i + \alpha \sqrt{\gamma}( {T}^{k}_i - \gamma_{ij} {\tau}^{jk}_{T})] = \sqrt{\gamma} {{R^S}}_i~,
\label{evol_S_sgs}  \nonumber \\
&& \partial_t (\sqrt{\gamma} {U}) + \partial_k [- \beta^k \sqrt{\gamma} {{U}} + \alpha  \sqrt{\gamma} {S}^{k}]  = \sqrt{\gamma} {R^U}  ~,
\label{evol_U_sgs}  \nonumber \\
&& \partial_t (\sqrt{\gamma} {B}^i) + \partial_k [\sqrt{\gamma}(- \beta^k {{B}}^i  +  \beta^i {{B}}^k) 
\nonumber \\
&& \quad\quad\quad\quad + \alpha \sqrt{\gamma} ({\gamma}^{ki} {{\phi}} + {M}^{ki} - {\tau}^{ki}_{M} )] = \sqrt{\gamma} {{R_B}}^i ~,
\label{evol_B_sgs} \nonumber \\
&& \partial_t (\sqrt{\gamma} {{\phi}}) + \partial_k [- \beta^k \sqrt{\gamma} {{\phi}} + \alpha\, c_h^2 \sqrt{\gamma}{{B}}^k] = \sqrt{\gamma} {R^{\phi}}~. 
\label{eq:evolution_sgs} 
\end{eqnarray}
The fluxes consist of the following standard terms:
\begin{eqnarray}
N^k &=& {v}^k {D} ~,\\
M^{ki} &=& B^{i} v^{k} - B^{k} v^{i} ~,  \\
T^{ki} &=& h W^2 v^k v^i - E^k E^i - B^k B^i  + \gamma^{ki} \left[ p+ \frac{1}{2}(E^2 + B^2 ) \right]  \nonumber \\
&=& \frac{1}{2} \left({v}^i {S}^j + {v}^j {S}^i \right) +  \gamma^{ij} {p} - \frac{1}{{W}^2} \bigg( {B}^i {B}^j - \frac{1}{2} \gamma^{ij} {B}^2  \bigg) \nonumber \\    
&-& \frac{1}{2} ({B}^k {v}_k) \bigg[ {B}^i {v}^j + {B}^j {v}^i - \gamma^{ij} ({B}^m {v}_m)  \bigg]~,
\end{eqnarray}
(where $E^i=-\epsilon^{ijk}v_j B_k$), and of the additional SGS terms:
\begin{eqnarray}
\tau^{k}_{N}  = -~{\cal C_N}~\xi \, H_{N}^k ~~, \nonumber \\
\tau^{ki}_{T} = -~{\cal C_T}~\xi \, H_{T}^{ki} ~~, \nonumber \\
\tau^{ki}_{M} = -~{\cal C_M}~\xi \, H_{M}^{ki} ~~. \label{eq:sgs_gradient}
\end{eqnarray}
The cumbersome expressions of the tensors $H$ have been obtained in detail for the special \cite{carrasco19} and general relativistic \cite{vigano20} cases. Here we apply the latter, following the expressions reported in the Appendix~\ref{app:sgs_forms}. The coefficient $\xi= \gamma^{1/3} \Delta^2/24$ has the proportionality to the spatial grid squared, which is typical of SGS models and ensures by construction the convergence to the continuous limit (vanishing SGS terms for an infinite resolution). Importantly, for each equation there is a pre-coefficient ${\cal C}_i$, which is meant to be of order one for a numerical scheme having a mathematically ideal Gaussian filter kernel and neglecting higher-order corrections. However, finite-difference numerical methods are usually more dissipative (and dispersive). Therefore, as shown in \cite{vigano19b,carrasco19,vigano20}, the value that best mimics the feedback of small scales onto the large scales in a LES can differ depending partially on the numerical methods employed and on the specific problem. In practice, one needs a calibration of the different SGS parameters to maximize the effectiveness of the gradient model.

Finally, the set of source terms in \eqref{eq:evolution_sgs}, $\{R^U,R^S_i,R_B^i,R^\phi\}$, written already as a function of conformal variables, can be found explicitly in \cite{vigano20}.
SGS terms are applied to the fluid equation only, considering the full general relativistic setting, with the assumption that the metric components are smooth and slowly varying, as compared to the turbulent and shocked matter fields (see \cite{vigano20} for a discussion).

\section{Numerical setup}\label{sec:setup}

\subsection{Numerical methods}

As in our previous works, we use the code {\sc MHDuet}, generated by the platform {\sc Simflowny} \cite{arbona13,arbona18} and based on the {\sc SAMRAI} infrastructure \cite{hornung02,gunney16}, which provides the parallelization and the mesh refinement. The code has been deeply tested for different scenarios \cite{palenzuela18,vigano19,vigano20,liebling20}, including basic tests of MHD and GR. Briefly, it uses: fourth-order-accurate operators for the spatial derivatives in the SGS terms and in the Einstein equations (the latter are supplemented with sixth-order Kreiss-Oliger dissipation); a high-resolution shock-capturing method for the fluid, based on the  Lax-Friedrich flux splitting formula \cite{shu98} and the fifth-order reconstruction method MP5 \cite{suresh97}; a fourth-order Runge-Kutta scheme with a small enough time step $\Delta t \leq 0.4. ~\Delta$; and an efficient and accurate treatment of the refinement boundaries when sub-cycling in time~\cite{McCorquodale:2011,Mongwane:2015}.  A complete assessment of the implemented numerical methods can be found in \cite{palenzuela18,vigano19}.

\begin{table*}[ht]
	\begin{tabular}{ |c|c|c|c|c|c| }			
		\hline
		Case
		& ${\cal C_M}$
		& ${\cal C_T}={\cal C_N}$
		& Refinement levels
		& Domain 
		of finest grid (km)
		& Finest $\Delta$ (m)
		\\ \hline
		{\tt C0 LR} & 0 & 0 & 5 FMR &  [-35,35] & 147 \\
		{\tt C0 MR} & 0 & 0 & 5 FMR+1 AMR & [-18,18] & 74 \\
		{\tt C0 HR} & 0 & 0 & 5 FMR+2 AMR & [-9,9] & 37 \\
		{\tt CM8} & 8 & 0 & 5 FMR & [-35,35] &  147 \\
		{\tt CM8C1} & 8 & 1 & 5 FMR & [-35,35] & 147 \\
		{\tt CM8C2} & 8 & 2 & 5 FMR & [-35,35] & 147 \\
		{\tt CM8C4} & 8 & 4 & 5 FMR & [-35,35] & 147 \\		
		{\tt C8} & 8 & 8 & 5 FMR & [-35,35] & 147 \\
		\hline
	\end{tabular}
	\caption{{\em Parameters of the simulations:} different resolutions, mesh refinement setup (with the finest grid spacing $\Delta$) and values of ${\cal C}_i$. Each setup is adopted at the merger time, while the inspiral phase is common to all of them and is run under the {\tt C0 LR} configuration. The domain of the finest AMR grid for the {\tt MR} and {\tt HR} cases changes with time, so that the values here indicated are only approximated.}
	\label{tab:models}
\end{table*}

\subsection{EoS and conversion to primitive variables}

We consider a hybrid EoS during the evolution, with two contribution to the pressure. On one side, we use the piecewise polytrope fit to the SLy zero-temperature EoS~\cite{read09}, defined by $p=K_i \rho^ {\Gamma_i}$, where $i=0,1,2,3$ indicates each of the four segments delineated by the transition density values $\log \rho = \{14.165,14.7,15.0\}$, $\Gamma_i = \{1.35692,3.005,2.988,2.851\}$ and $K_0 = 3.59389\times 10^{13}$ (all in cgs units).
On the other hand, thermal effects are modeled by an additional pressure contribution given by the ideal gas EoS, with adiabatic index $\Gamma_{\rm th}= 1.75$ \cite{bauswein10}. 

The conversion from the evolved or conserved fields to the primitive or physical ones is performed by using the procedure described in our previous works \cite{vigano20,liebling20}. An exception is the highest resolution simulation, for which the strong magnetic fields developed in low-density regions forced us to use a more robust procedure \cite{kastaun20}. To minimize further failures on the recovery procedure outside the dense regions, we impose a minimum density of $6.1 \times 10^{7}~\rm{g~cm^{-3}}$, with the regions having such values referred hereafter as atmosphere. Moreover, we apply the SGS terms only in regions where the density is higher than $6.1 \times 10^{11}~\rm{g~cm^{-3}}$ in order to avoid spurious effects near the stellar surface. Since the remnant's maximum density is above
$10^{15} ~\rm{g~cm^{-3}}$, the SGS model is accounted for only in the most dense regions of the star.

\subsection{Initial conditions}

The initial data is created with the {\sc Lorene} package~\cite{lorene}, using the same piecewise polytropic EoS described above. We consider an equal-mass BNS in quasi-circular orbit, with an irrotational configuration having a separation of $37.7$ km and an angular frequency of $2254\ \rm{rad~s^{-1}}$. The total mass of the system is $M=2.67~M_{\odot}$.

Each star initially has a purely poloidal magnetic field confined in its interior, calculated from a vector potential 
$A_ {\phi} \propto r^2 (P - P_{cut})$, where $P_{cut}$ is a hundred times the pressure of the atmosphere and $r$ the distance to the axis perpendicular to the orbital plane passing through the centre of each star. The maximum magnetic intensity (at the centres) is $5\times 10^{11}$ G, orders of magnitude lower than the large initial fields of other simulations (e.g., \cite{kiuchi15,ruiz16,kiuchi18,ciolfi2019,ciolfi2020collimated,ruiz2020}) and compatible with the upper range of the expected realistic intensities for old NSs. Such values are also at the lower border of the computational feasibility, since the accurate evolution for too small ratios of magnetic-to-kinetic pressure is hampered by round-off errors.

\section{Results}\label{sec:results}

\begin{figure*}
	\centering
	\includegraphics[width=0.36\textwidth]{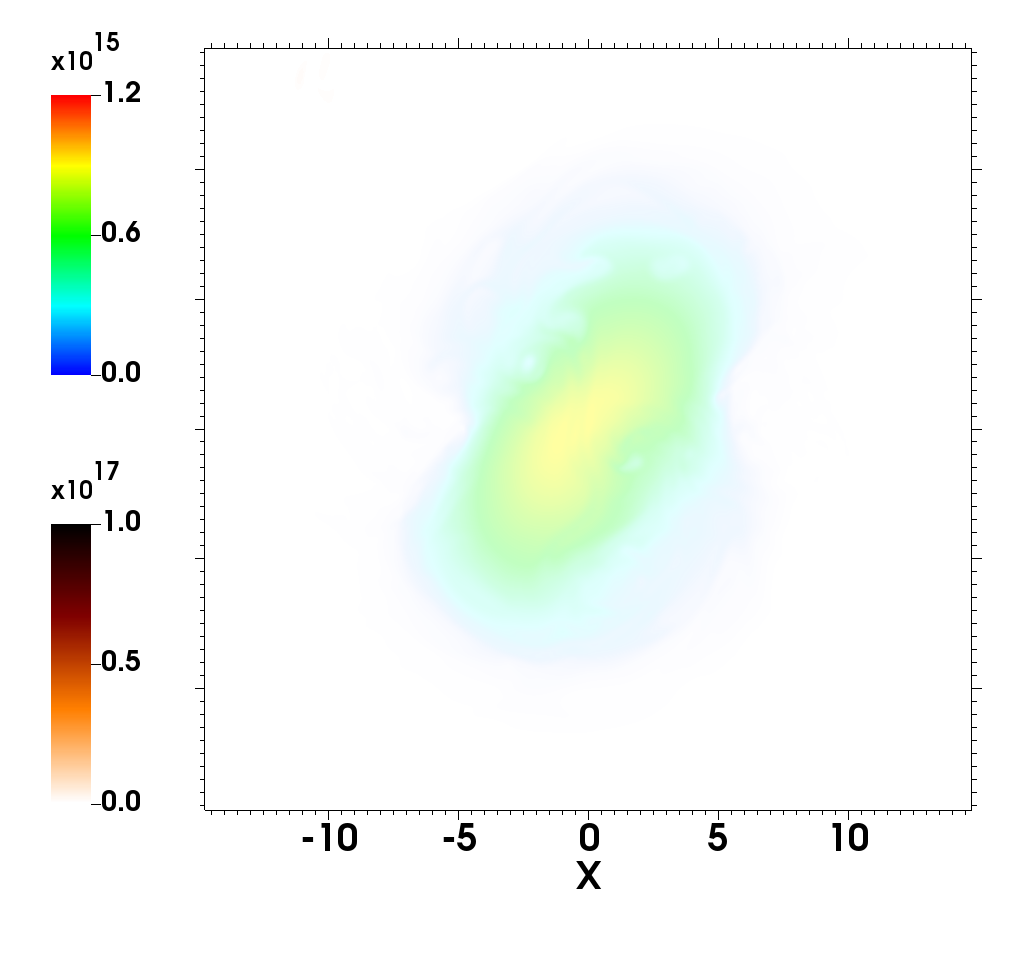}
	\includegraphics[width=0.3\textwidth,trim=170 0 0 0,clip]{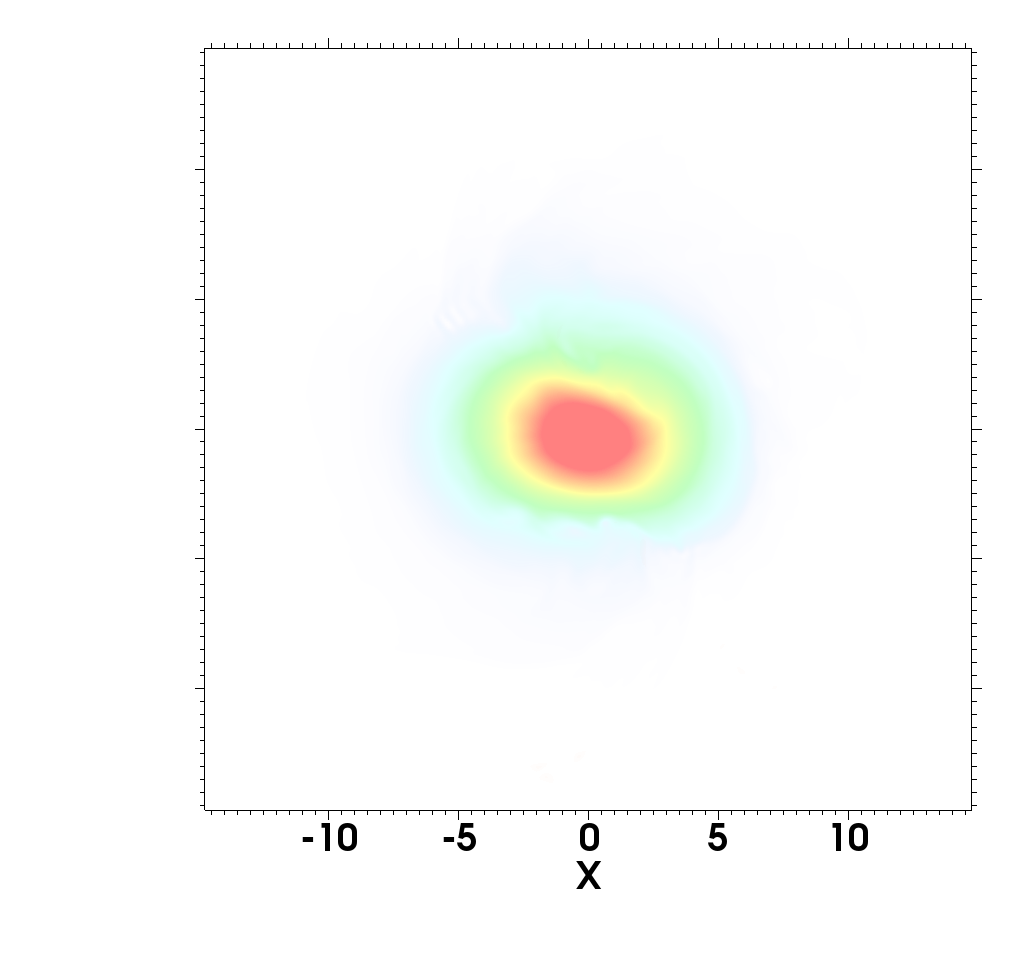}
	\includegraphics[width=0.3\textwidth,trim=170 0 0 0,clip]{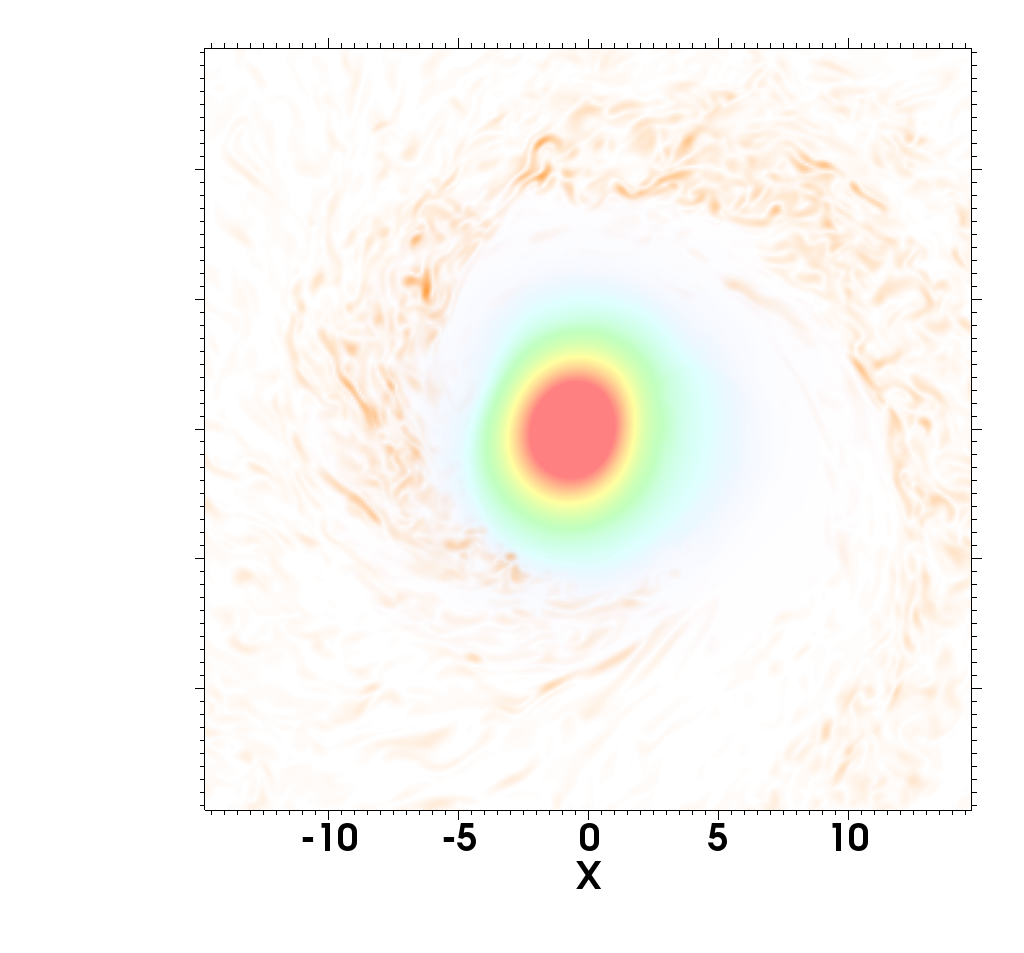} \\
	\includegraphics[width=0.36\textwidth]{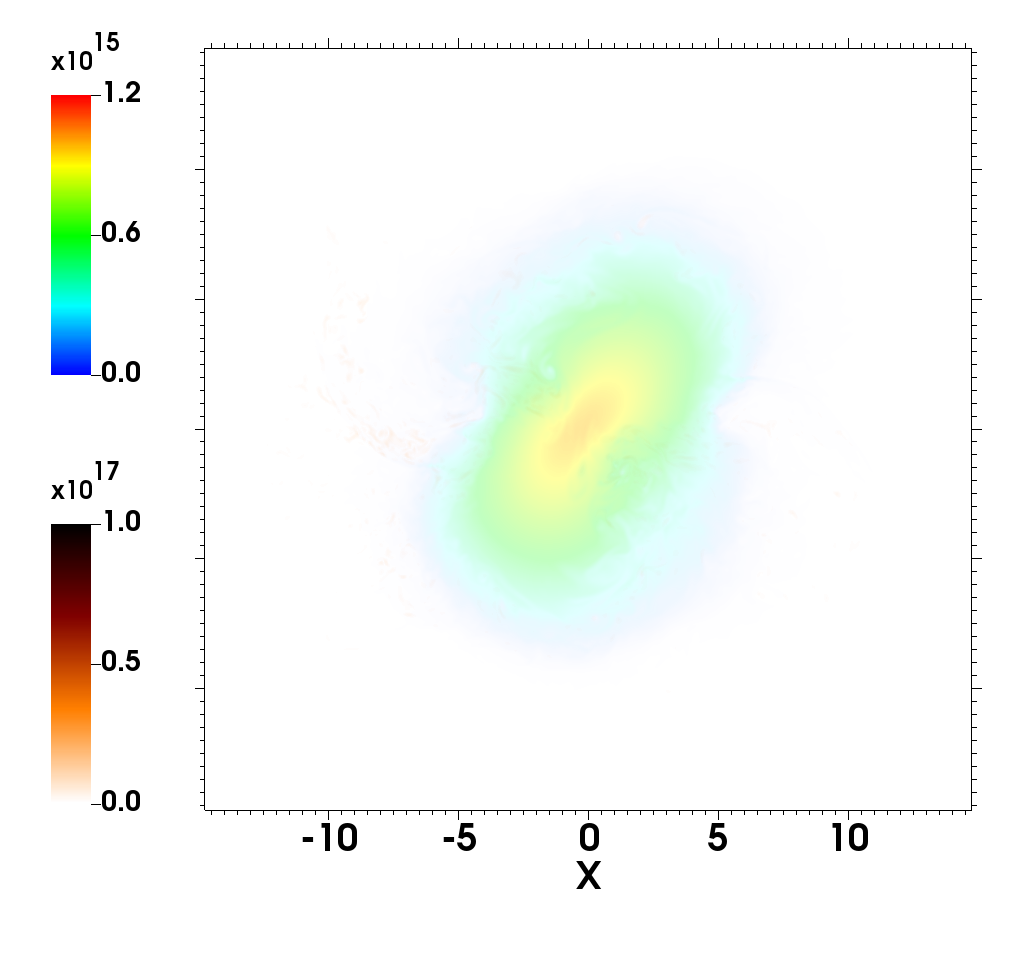}
	\includegraphics[width=0.3\textwidth,trim=170 0 0 0,clip]{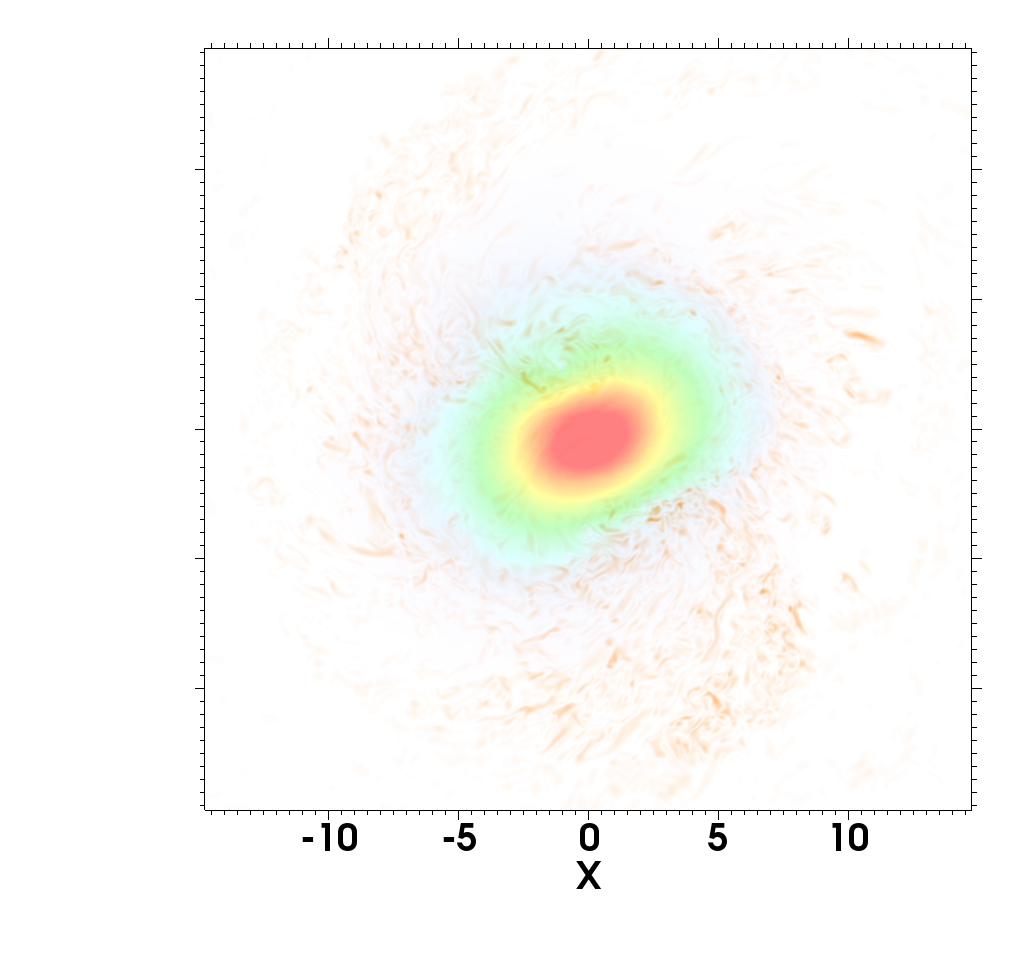}
	\includegraphics[width=0.3\textwidth,trim=170 0 0 0,clip]{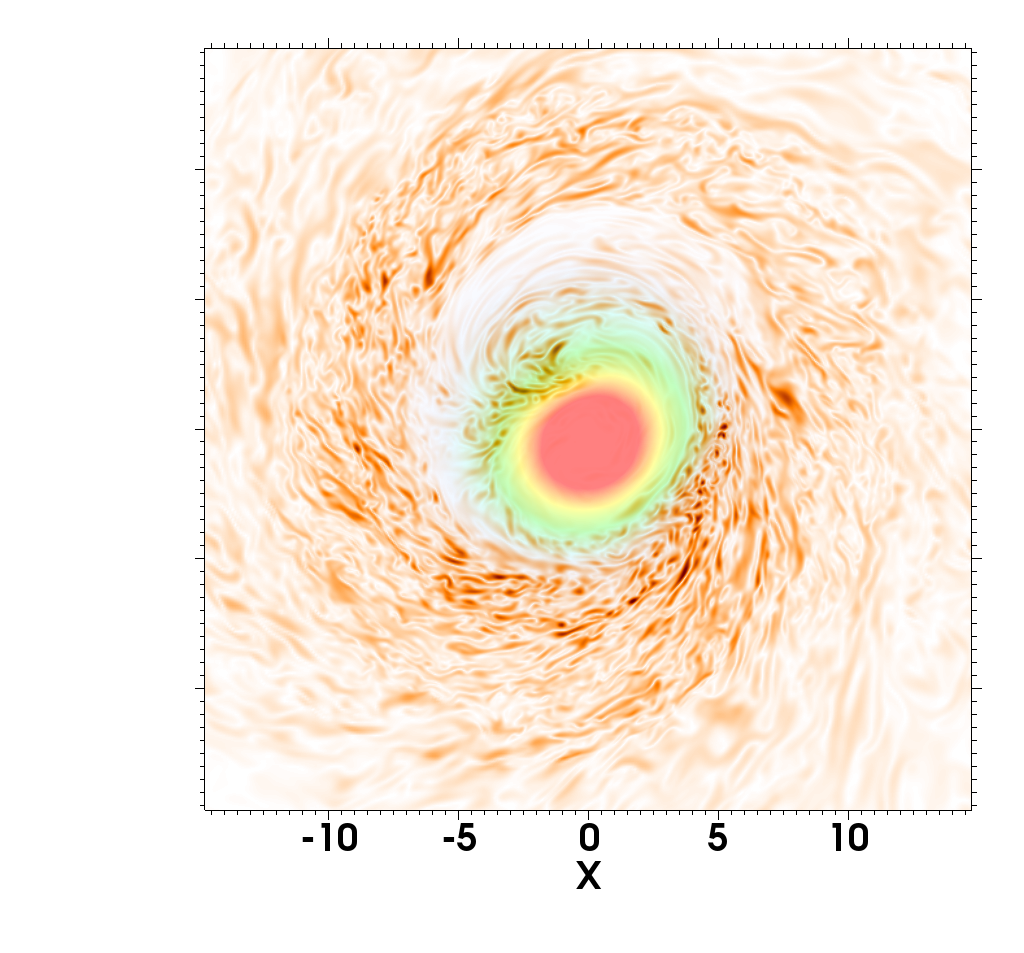} \\	
	\includegraphics[width=0.36\textwidth]{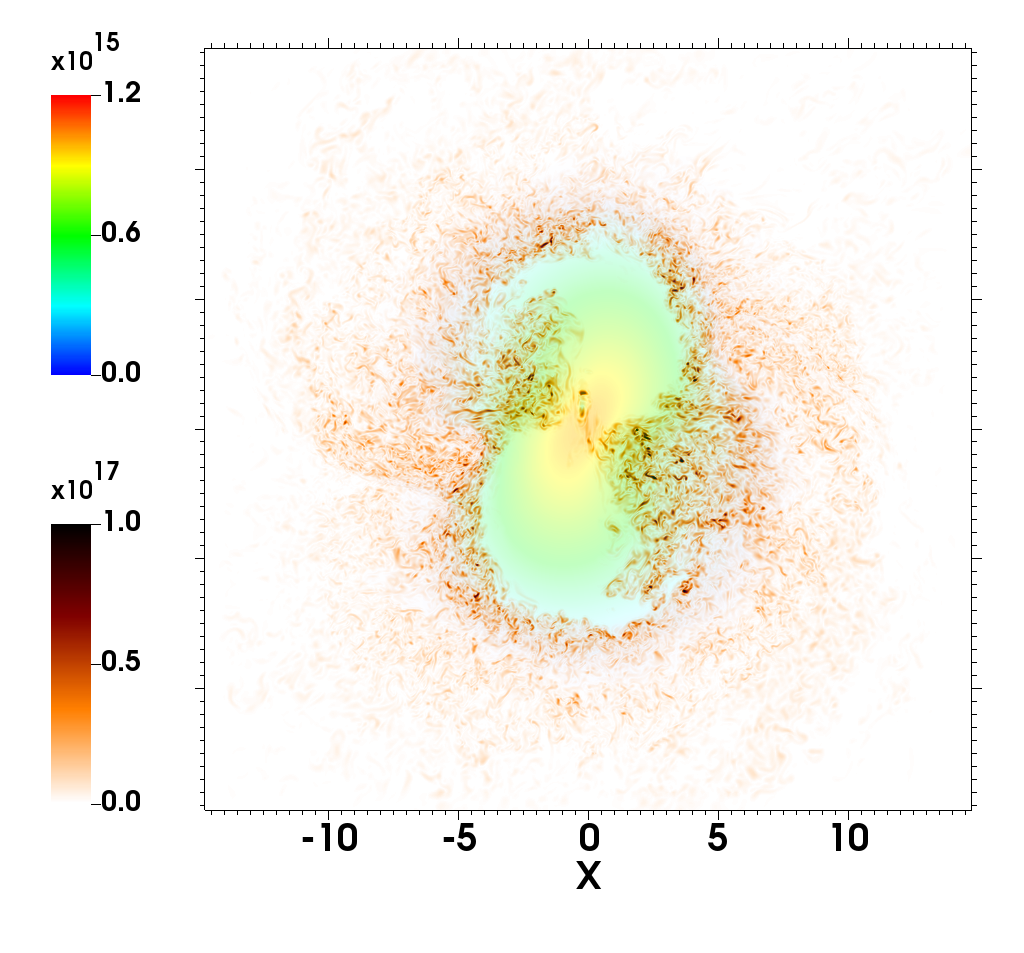}
	\includegraphics[width=0.3\textwidth,trim=170 0 0 0,clip]{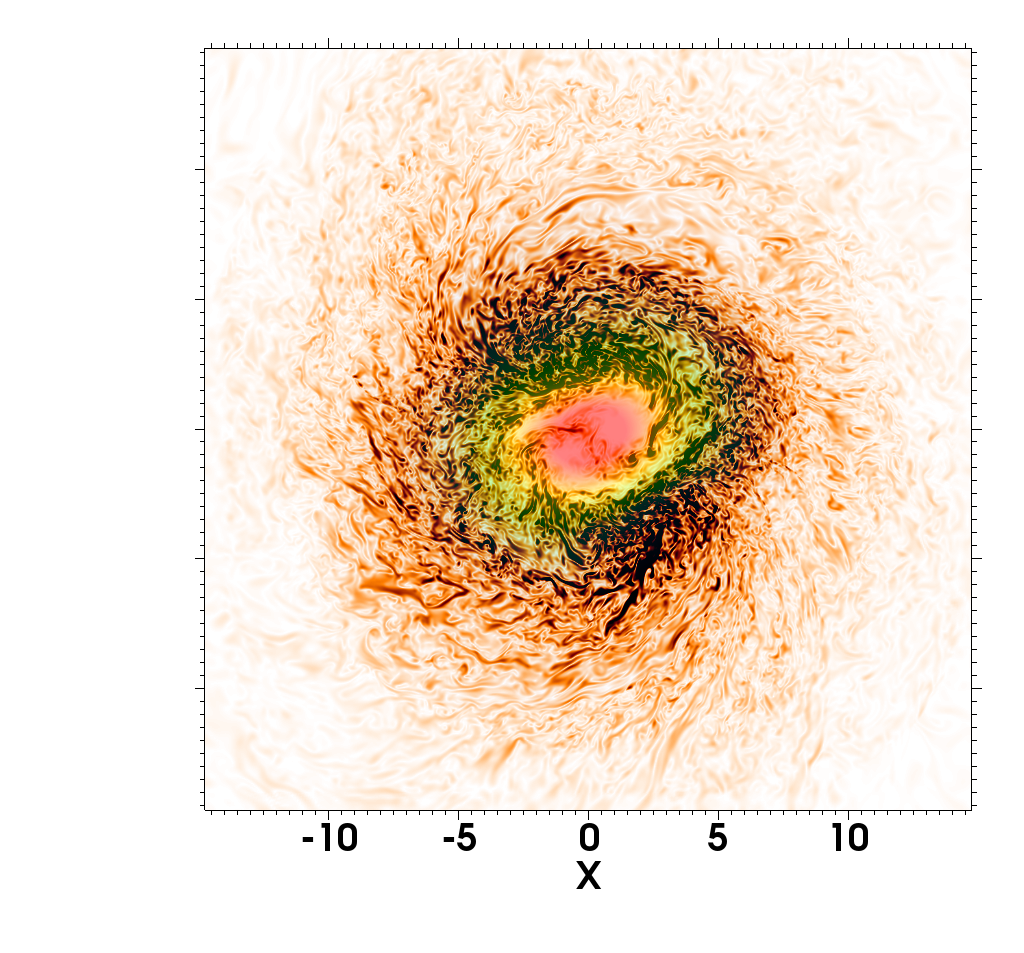}
	\includegraphics[width=0.3\textwidth,trim=170 0 0 0,clip]{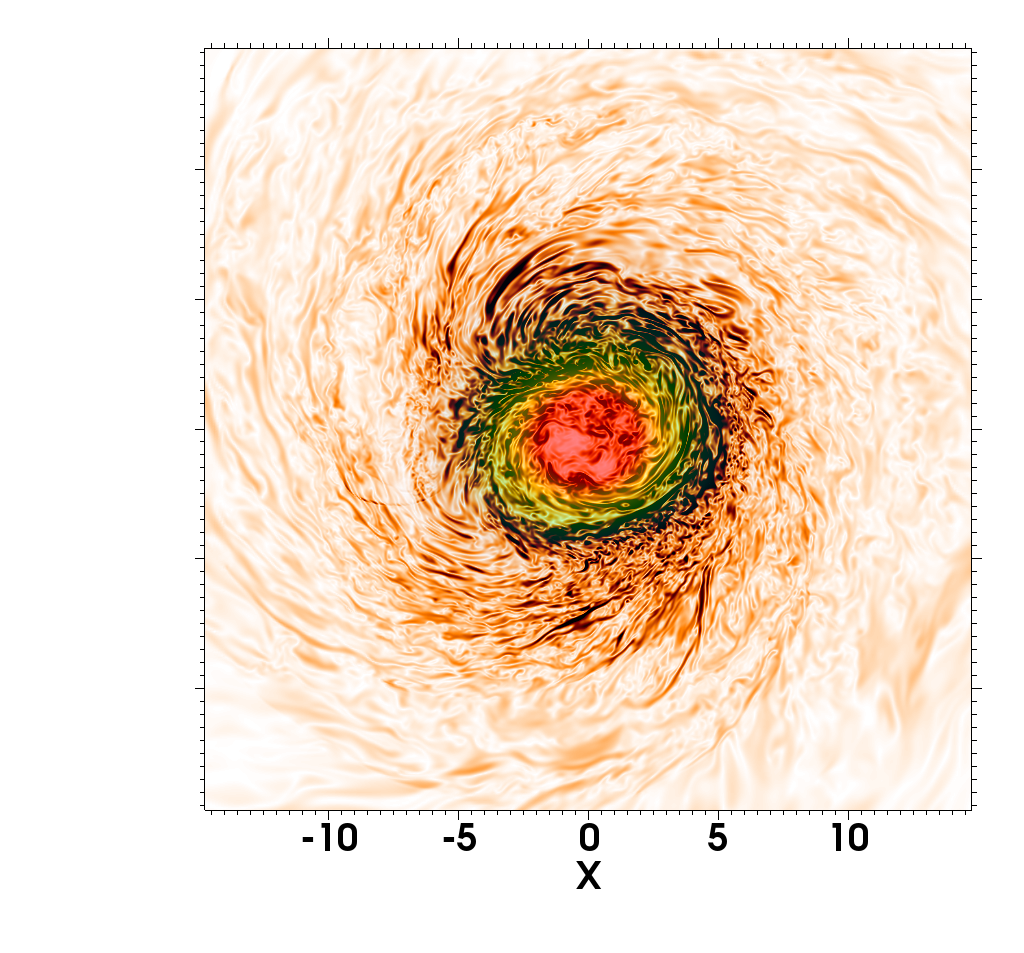}	
	\caption{{\em iLES of the BNS merger with three different resolutions.} Evolution of the solution on the orbital plane for {\tt LR} (top), {\tt MR} (middle) and {\tt HR} (bottom) at $t=2.5\ ms$ (left), $t=5\ ms$ (centre) and $t=10\ ms$ (right) after the merger. The rainbow and brownish colour scales represent the values of density and magnetic field in cgs units, while the length is given in geometrical units (corresponding to $1.47\ km$).}
	\label{fig:slice_magnetic_iLES}
\end{figure*}

\begin{figure*}
	\centering
	\includegraphics[width=0.35\textwidth]{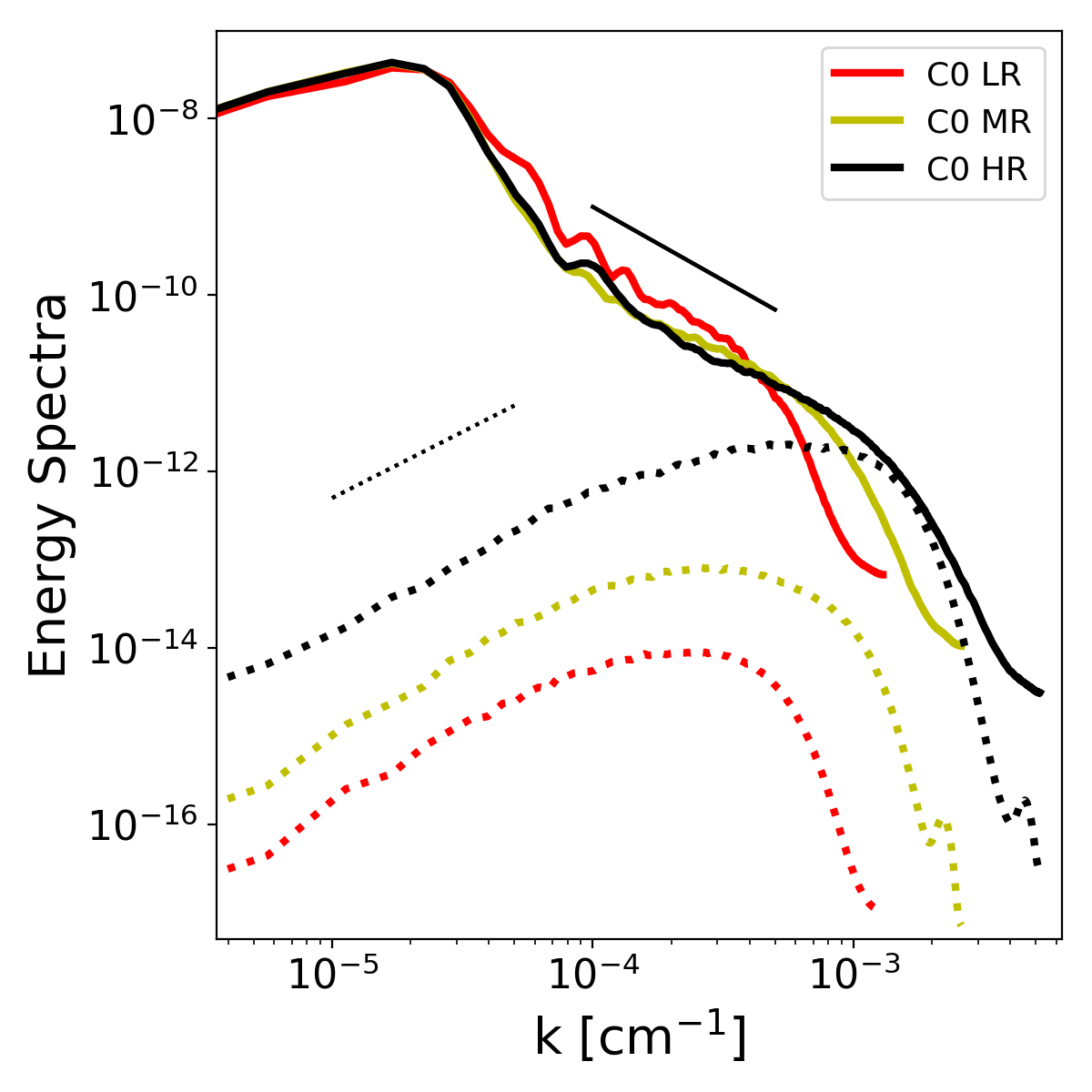}	\includegraphics[width=0.35\textwidth]{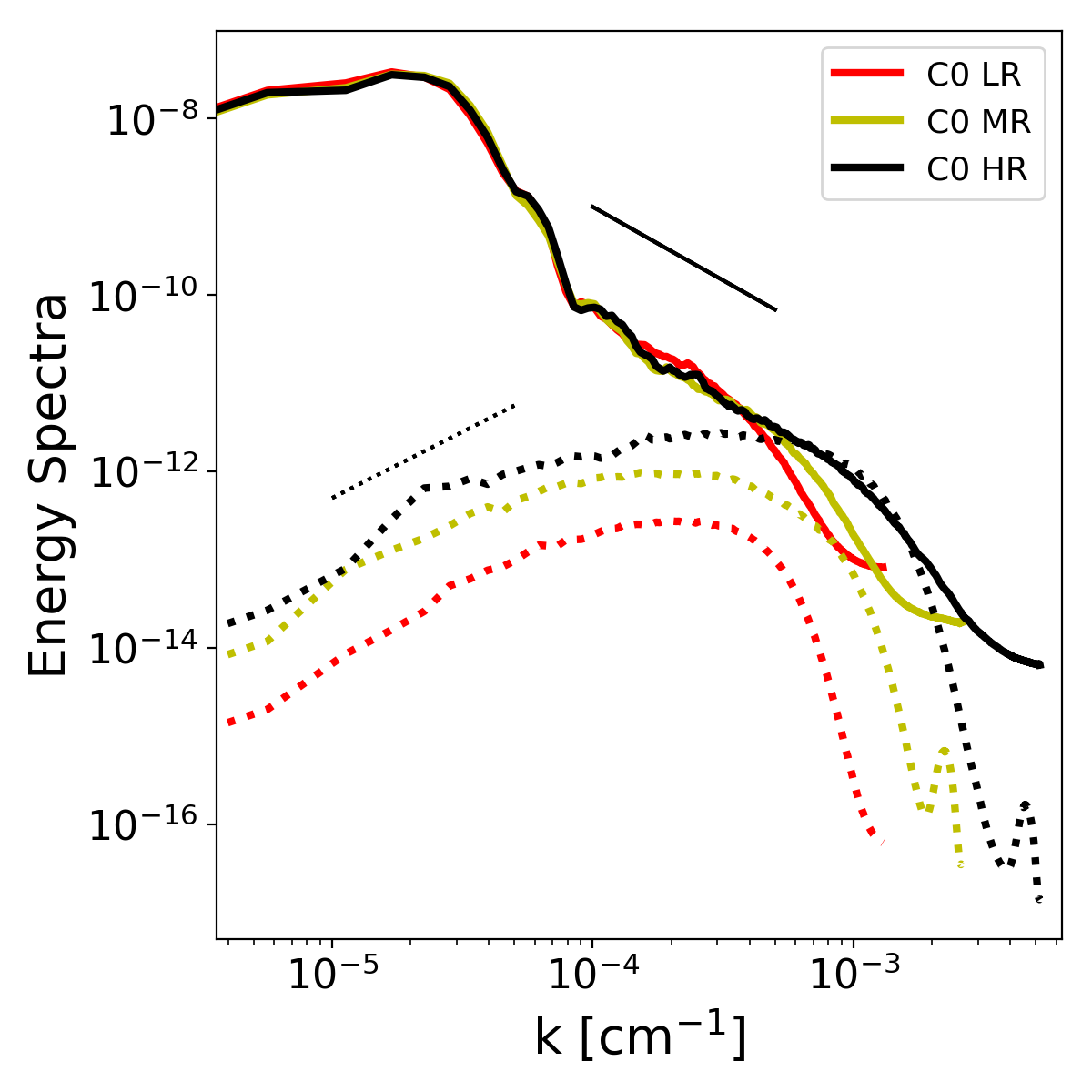} 
	\caption{{\em Energy spectra for simulations with different resolutions.} Kinetic (solid line) and magnetic (dashed line) spectral energy distributions as function of the angular wave number, for the three different resolutions, at $t=5\ ms$ (left) and $t=10\ ms$ (right). The solid and dotted black lines represent the Kolmogorov ($k^{-5/3}$) and Kazantsev ($k^{3/2}$) slopes, respectively. The energy spectra hereafter are in arbitrary units.}
	\label{fig:spectra_iLES}
\end{figure*}

We consider a numerical cubic domain, ranging from $[-384,384]$ km along each direction, large enough to reduce contamination from the boundaries. Our initial binary system evolves for 2-3 orbits before merging and forming a differentially rotating remnant that relaxes to an hypermassive-neutron star (HMNS) in a few milliseconds. We follow such inspiral with five nested levels of Fixed Mesh Refinement (FMR), each being a cube doubling the resolution of the previous one. The smallest and finest of them is $70$ km wide, thus it encloses the stars during the inspiral and the forming remnant. At the merger time (hereafter, $t=0$), we have then considered different simulations, summarized in Table~\ref{tab:models}. 

First, we present standard simulation without SGS terms (${\cal C}_i=0$), also called implicit LES\footnote{The definition of an iLES applies actually to any standard simulation and comes from the fact that any numerical scheme has some dissipative and dispersive character which implicitly enters in the discretized equations. However, such implicit SGS modeling is not trivial to be assessed and is virtually impossible to be controlled or calibrated.} (iLES, denoted by {\tt C0} hereafter), with grid spacing corresponding to low ({\tt LR}, finest level: $147$ m), medium ({\tt MR}, $74$ m) and high resolution ({\tt HR}, $37$ m). The {\tt LR} case has five FMR levels, like in the inspiral. In the {\tt MR} and {\tt HR} cases, we activate one and two additional Adaptive Mesh Refinement (AMR) levels (again doubling the resolution of the previous level), respectively, describing the regions exceeding certain density thresholds properly set, in order to better resolve the remnant.

Secondly, we perform LES with {\tt LR} including the SGS models. Here we report the cases with a fixed ${\cal C_M}=8$, spanning ${\cal C_T}={\cal C_N}=\{0,1,2,4,8\}$. Other combinations of parameters with ${\cal C_M}={\cal C_T}={\cal C_N} > 2$  have been tested, but they produced an excessive dissipation in the momentum equation, leading to unrealistic results. Also, we have considered the non-relativistic limit of the SGS term in the induction equation proposed in \cite{vigano20} (i.e. neglecting the $H_v^k$ contribution in Eq.\eqref{Tauv}). In contrast to the box simulations results in \cite{vigano20}, we see that the relativistic corrections on these SGS terms produce here a significant increase of the magnetic field amplification, so we have kept the full expressions for our simulations.
From now on, we will refer to the LES simulations by labeling them in a schematic way according to the $\mathcal{C}_i$ values, as indicated in Table \ref{tab:models}.

\subsection{Results with different resolutions}

First, we consider the three iLES cases. In Fig.~\ref{fig:slice_magnetic_iLES} we show the density and the magnitude of the magnetic field, in the equatorial plane $z=0$, for the three resolutions (different rows) at $t=\{2.5,5,10\}$ ms (different columns) after the merger. In agreement with previous results \cite{kiuchi18}, the magnetic field grows on small structures especially in the outermost, less dense layers of the remnant, where plasma is closer to an equipartition between magnetic and kinetic energy. As expected, this amplification is enhanced by a finer grid, since smaller wavelengths grow faster in the KHI. At $t=2.5$ ms, the two cores are still clearly distinguishable, indicating that the remnant has not relaxed to a HMNS yet. At this early stage, magnetic fields locally exceed $\sim 10^{17}$ G only in the {\tt HR} simulation, with fine structures clearly visible. At $t=5$ ms, the remnant and surrounding disk are forming and turbulence drives the magnetic field amplification to maximum values of $\sim 5 \times 10^{17}$ G in the {\tt HR}, dropping one order of magnitude in the {\tt MR} simulation and another one for the {\tt LR} case (see a quantitative comparison of magnetic energy evolution below). It can be seen how the magnetic field is mostly confined to the outermost layers of the remnant, since the dense core is less prone to turbulent motions. At $t=10\,$ms, the strong magnetic fields has started to penetrate into the dense core of the remnant in the {\tt HR} run, while a significant  overall increase in the field strength is also noticeable for the lower resolutions. At this time, although small-scale structures still dominate, the rotation has acquired a visible imprint on the magnetic field distribution, developing spiral-like filaments at the outermost layers.

\begin{figure*}
	\centering
	\includegraphics[width=0.36\textwidth]{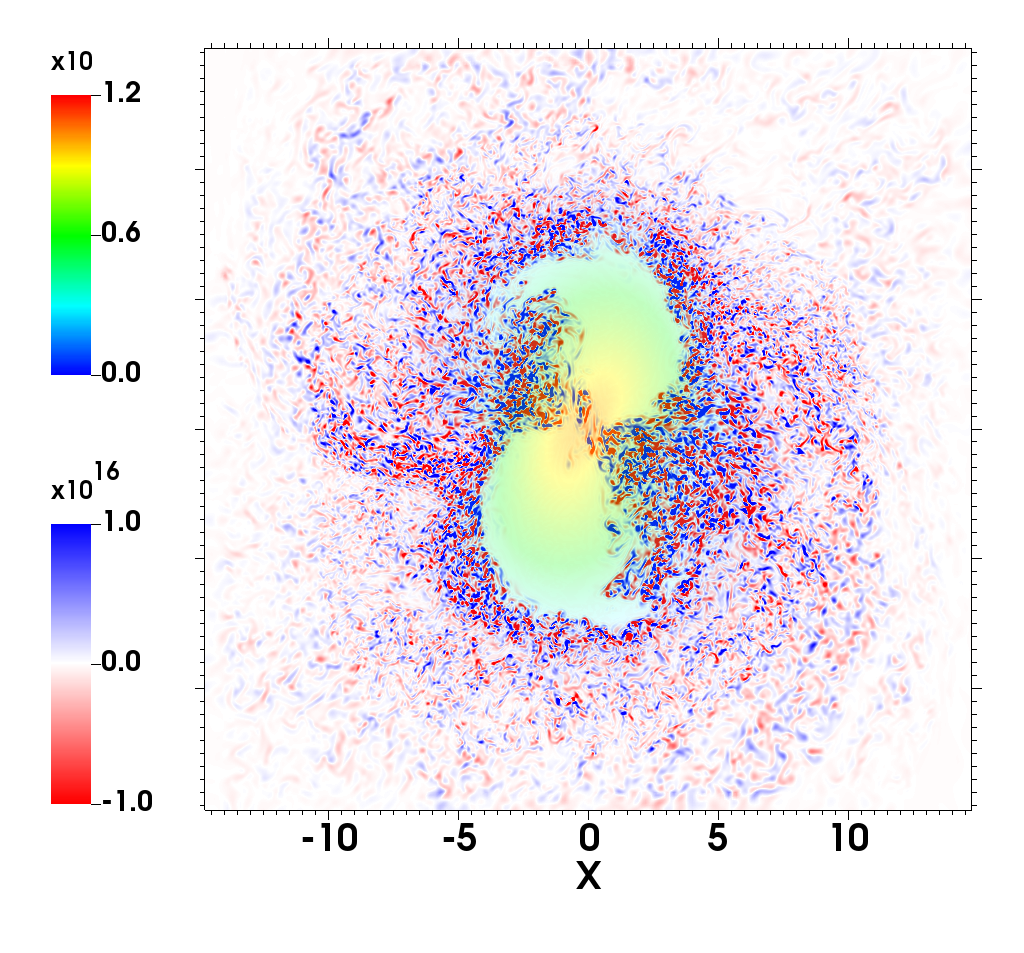}
	\includegraphics[width=0.3\textwidth,trim=170 0 0 0,clip]{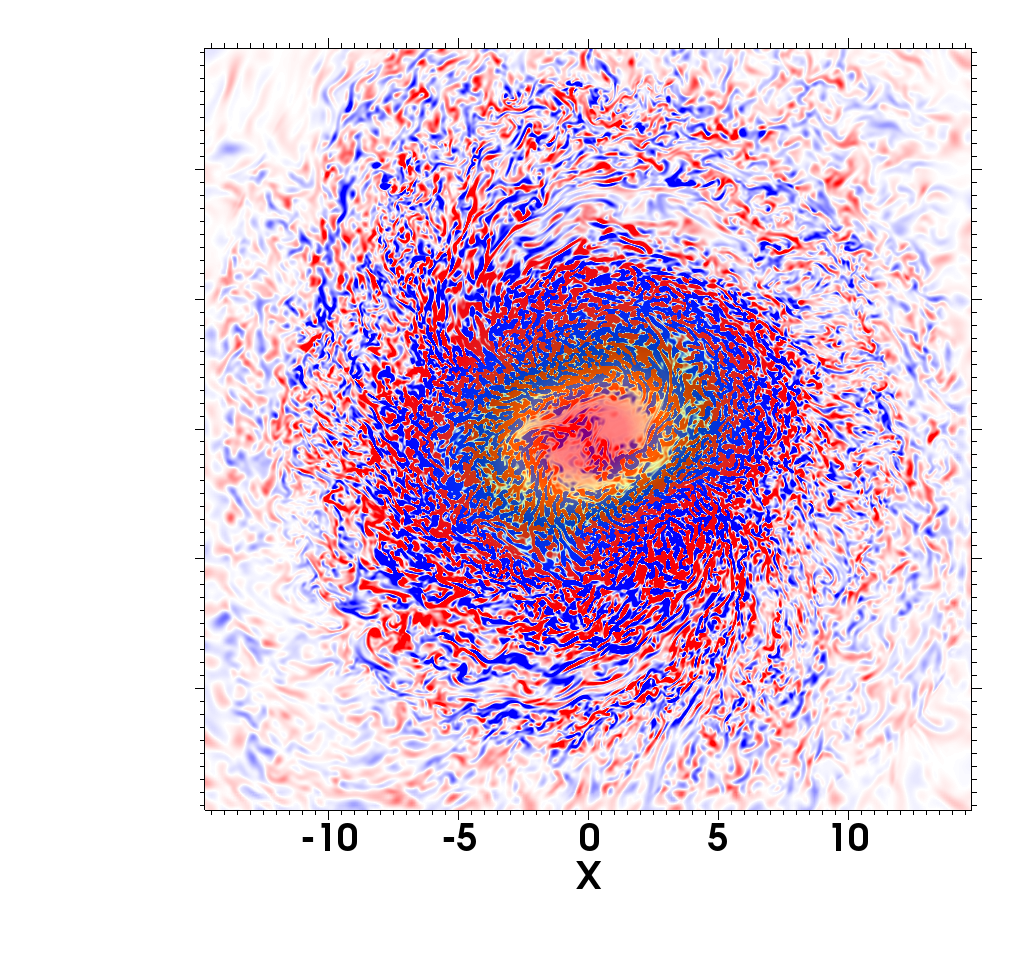}
	\includegraphics[width=0.3\textwidth,trim=170 0 0 0,clip]{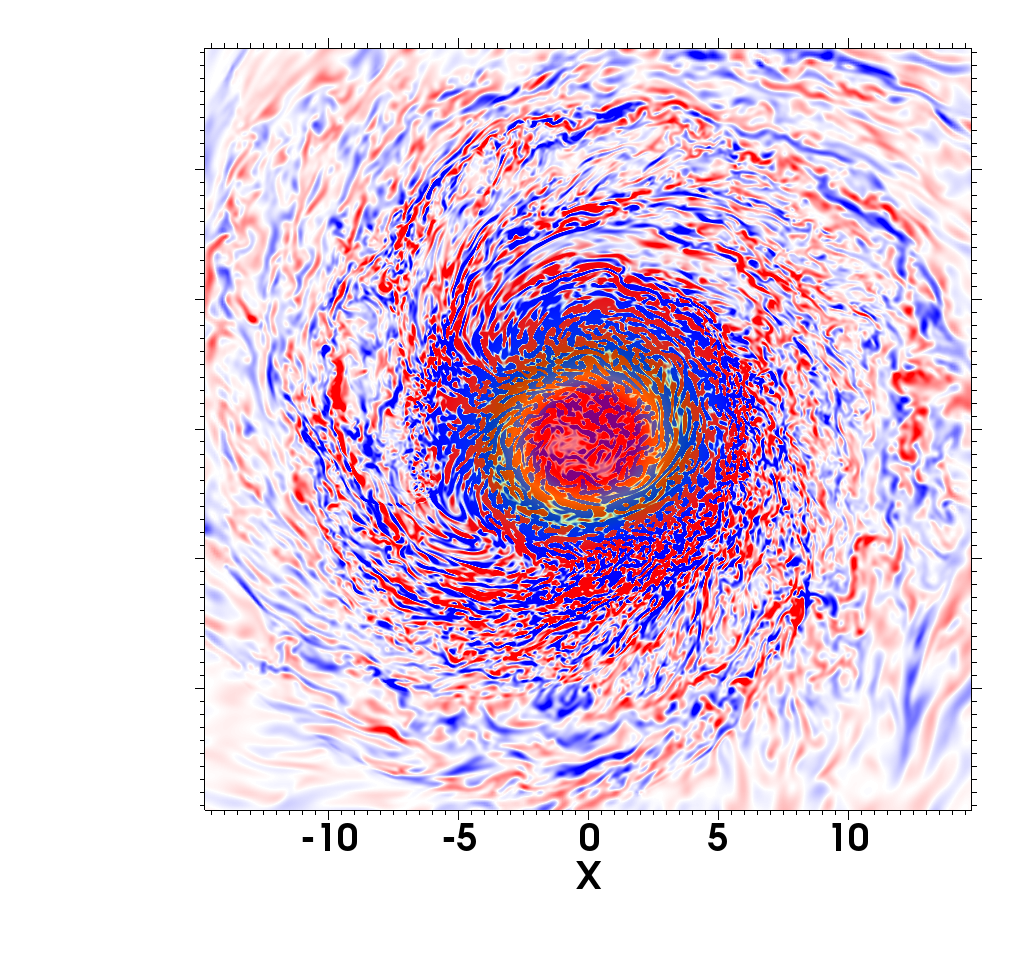} \\		
	\includegraphics[width=0.36\textwidth]{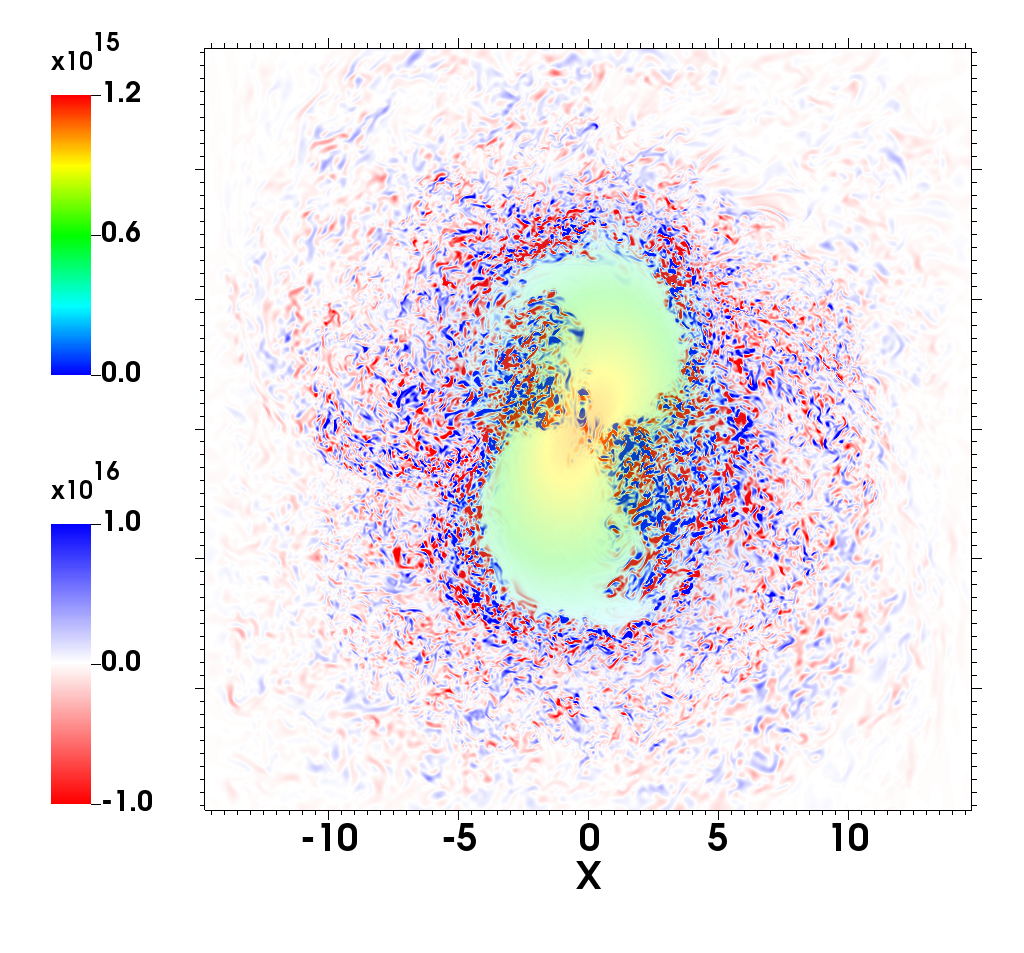}
	\includegraphics[width=0.3\textwidth,trim=170 0 0 0,clip]{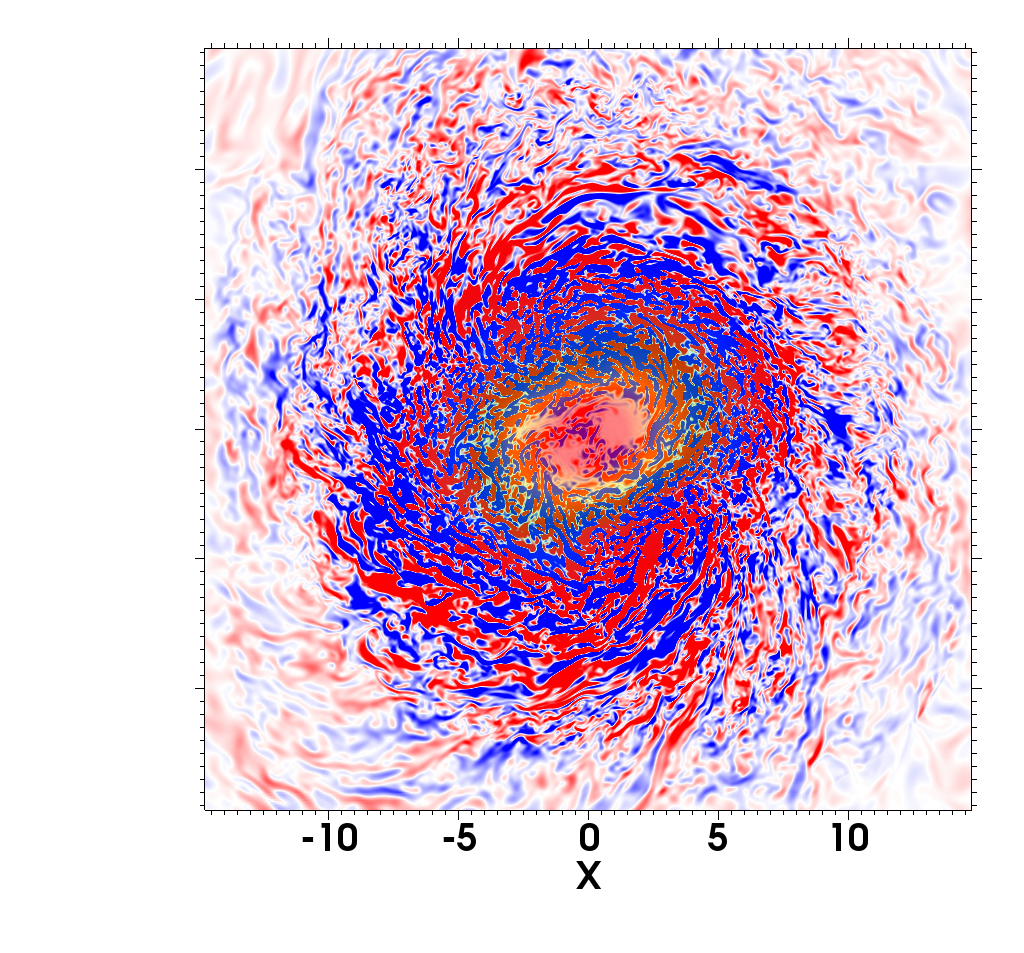}
	\includegraphics[width=0.3\textwidth,trim=170 0 0 0,clip]{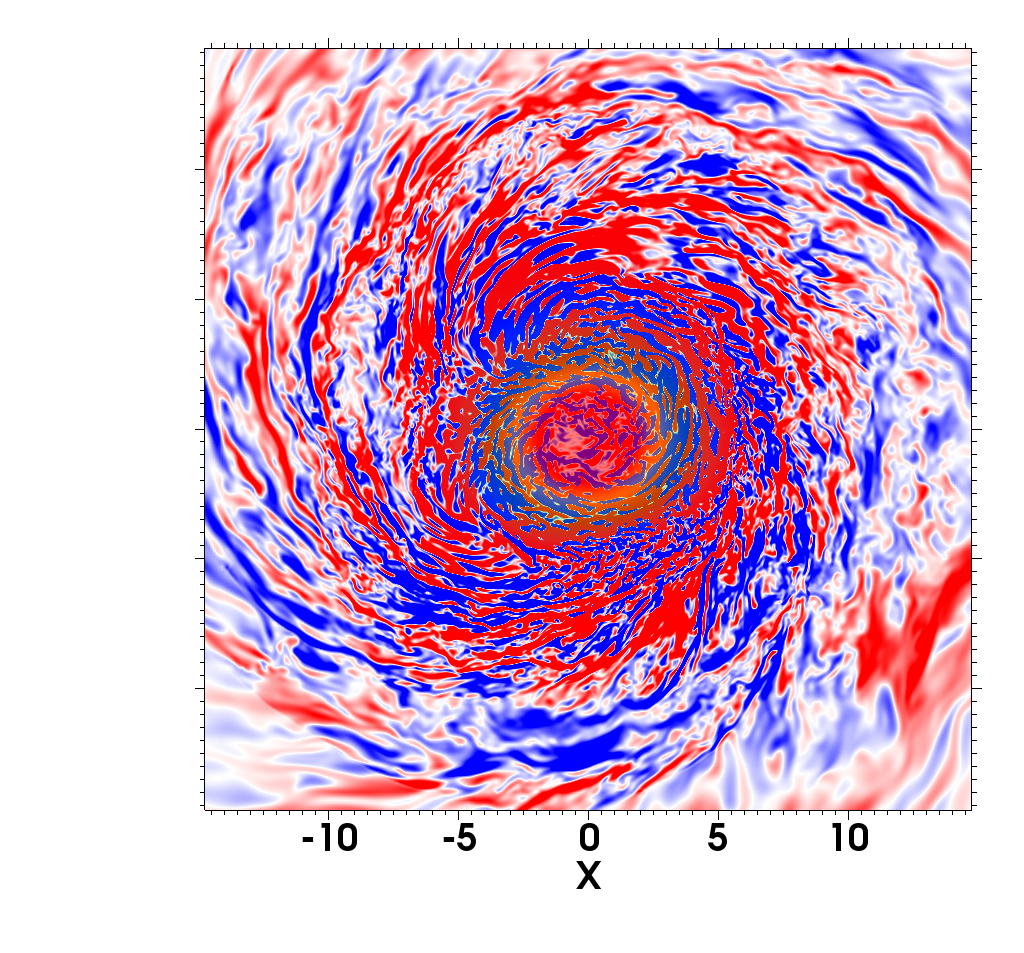} 
	\caption{{\em Evolution of magnetic field components.} Evolution of the {\tt HR} case of density and components of the magnetic field, indicated in blue-red scale (colours saturate at $\pm 10^{16}\ G$): the one perpendicular to the orbital plane (top) and the azimuthal one (bottom), at $t=2.5\ ms$ (left), $t=5\ ms$ (centre) and $t=10\ ms$ (right) after the merger.
	Units and colour scale of the density as in Fig.~\ref{fig:slice_magnetic_iLES}.
	}
	\label{fig:slice_magnetic_tor_L7}
\end{figure*}

\begin{figure*}[ht]
	\centering
	\includegraphics[width=0.35\textwidth]{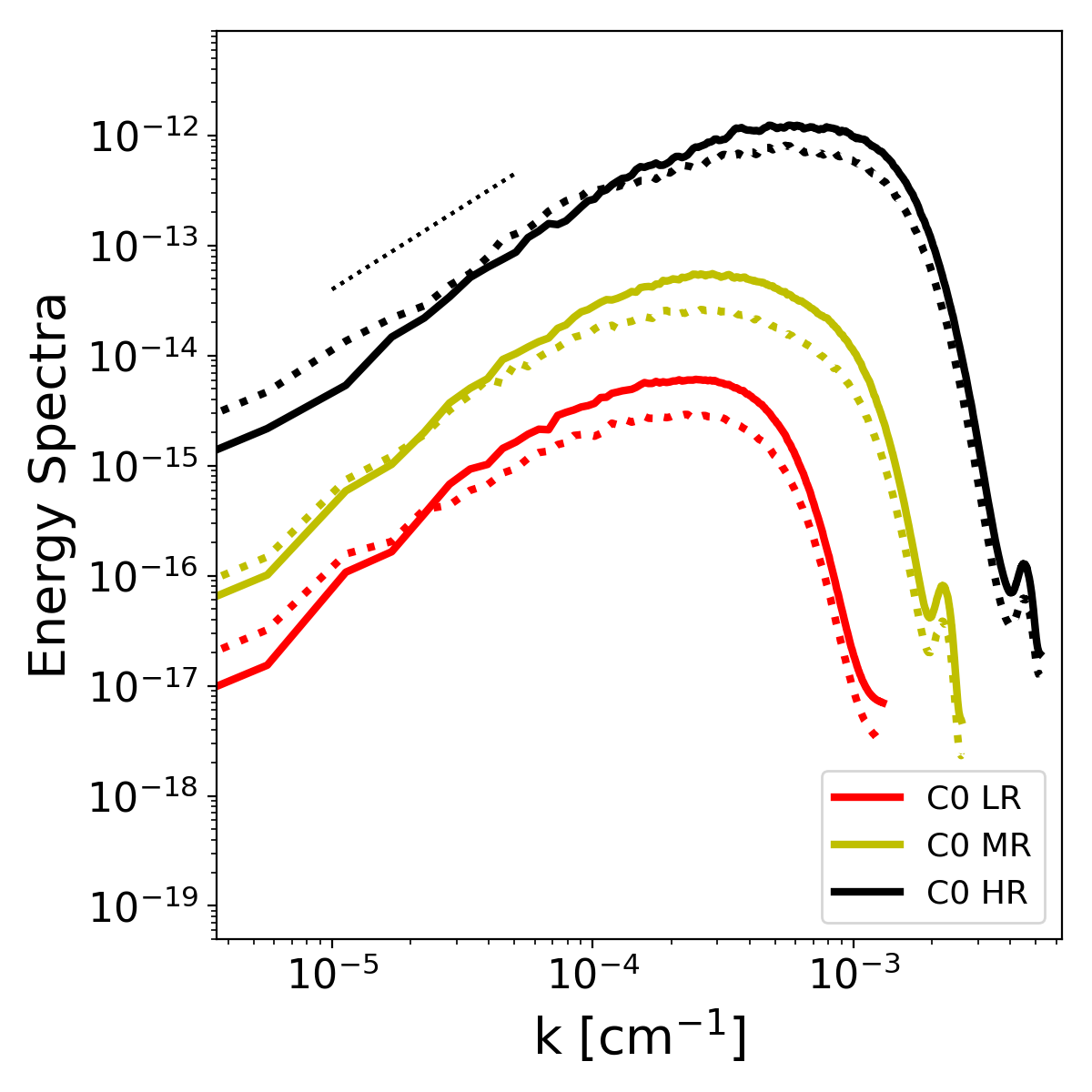}
	\includegraphics[width=0.35\textwidth]{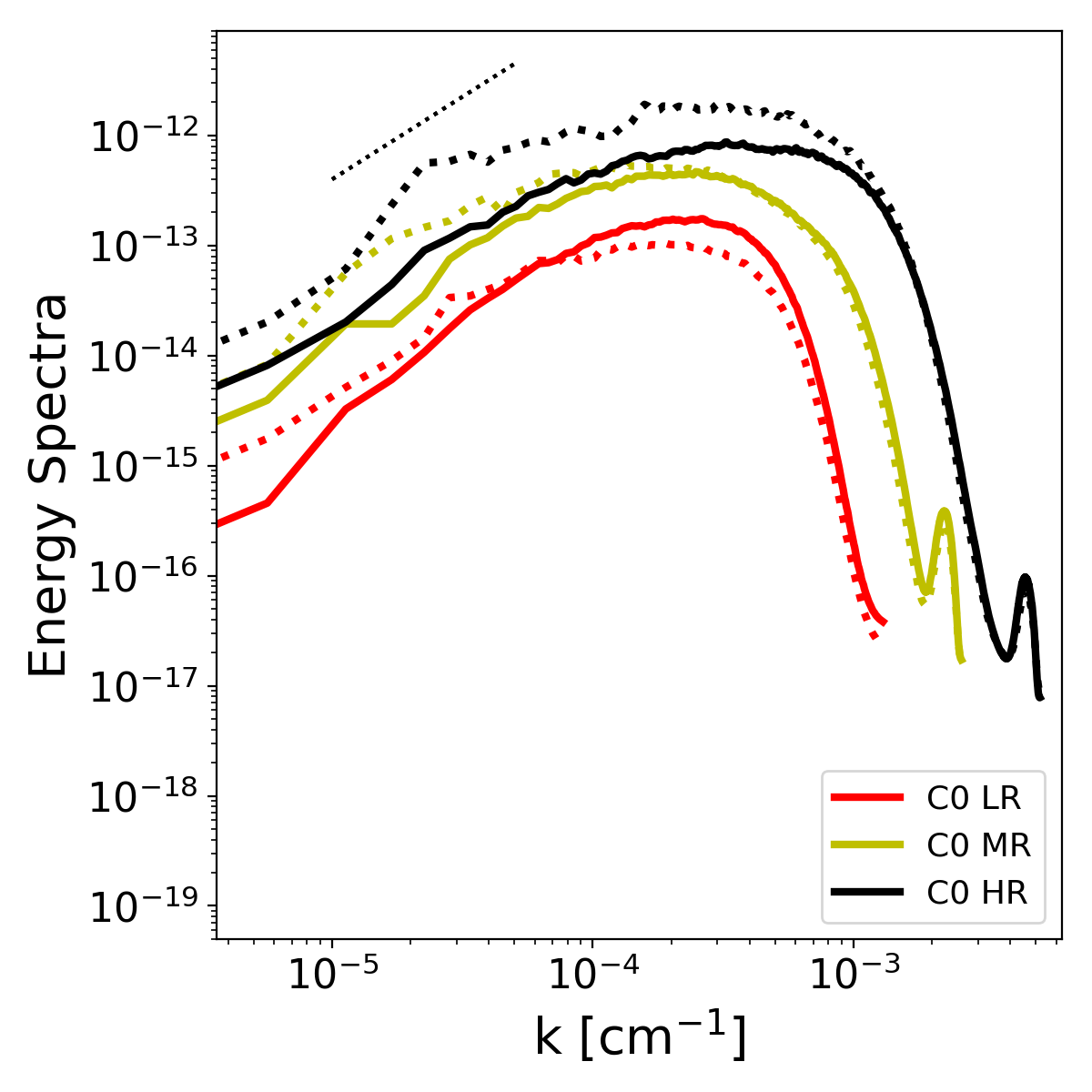}
	\caption{{\em Magnetic energy spectra by components.} Magnetic poloidal (solid line) and toroidal (dashed line) spectra of {\tt LR}, {\tt MR} and {\tt HR} cases at $t=5\ ms$ (left) and $t=10\ ms$ (right). The two components have similar profiles, although at $t=10\ ms$ the toroidal component is slightly larger than the poloidal one for the high resolution cases {\tt MR} and {\tt HR}.
	}
	\label{fig:spectra_tor_pol_iLES}
\end{figure*}

This visual inspection can be quantified by the study of the energy spectral distribution (see Appendix~\ref{app:spectra} for definitions and calculations), as shown in Fig.~\ref{fig:spectra_iLES} for $t=\{5,10\}$ ms.  Note that, in general, we can identify the inertial range  between scales much larger than $\Delta$ (around which numerical dissipation acts) and smaller than the energy-injection scales (set in this case by the rotation). In such range, the kinetic and magnetic spectra approximately follow the Kolmogorov ($k^{-5/3}$) and Kazantsev ($k^{3/2}$) slopes (dotted black lines in the figures hereafter), as expected in turbulent MHD scenarios.

The kinetic energy distribution is dominated by large scales, so that the resolution has a lower impact on it. Instead, the absence of a peak in the magnetic energy at low $k$ (at least until $10$ ms) means that there is no hint, at these times, for the creation of a strong, large-scale, ordered magnetic field. Small scales are the main form of magnetic energy storage, hence the importance of the numerical resolution. This can be clearly observed, especially at early times ($5$ ms, left panel): the higher the resolution, the larger the growth of the magnetic energy is, even though the spectra have the same profiles. At later times ($10$ ms, right panel), the difference between different resolutions greatly decreases, especially at large scales. Thus, pointing to a saturation of the KHI, achieved by all the three resolutions.

The magnetic amplification here illustrated presents the typical dynamical stages of the KHI, described e.g. in \cite{zrake13b} as follows: an initial \textit{startup} transient associated to the full development of the turbulent cascade (triggered at the merger); the \textit{kinematic phase}, in which the magnetic fields are still sub-dominant but grow exponentially, driven by an essentially hydrodynamical turbulent mechanism as in Kazantzev's theory \cite{kazantsev68}; 
the approach to \textit{saturation} when the magnetic field becomes strong enough as to back-react on the fluid motion and establish a dynamical balance signaled by kinetic/magnetic spectral equipartition at small scales. Generally speaking, the magnetic saturation levels are expected to converge (at least above certain threshold resolution \cite{zrake13b}), while the growth rates and timescales of each dynamical phase are highly sensitive to the numerical resolution.

Note that the drop in the spectra for high $k$ (approaching the upper limit set by $\pi/\Delta$) is due to the intrinsic numerical dissipation of the finite-volume scheme (spectral methods should not show it). Overall, the same behaviour was observed in our box simulations of the KHI \cite{vigano19b,vigano20}: a rising of the magnetic energy as smaller and smaller eddies develop, finally reaching equipartition of the spectral distribution at small scales (high $k$), while at large scales (low $k$) the kinetic energy always dominates.

It is also interesting to look separately at the magnetic components. In Fig.~\ref{fig:slice_magnetic_tor_L7}, we show the evolution of the component perpendicular to the orbital plane (top) and of the azimuthal one, again at $t=\{2.5,5,10\}$ ms. At the beginning, both components show very similar small structures, indicating a high degree of isotropy, proper of a developing turbulence that stretches and twists the initially weak large-scale magnetic seed. At later times, when differential rotation starts to dominate the kinematics, the magnetic structures tend to follow the rotation, partially losing the isotropy. In particular, at $t=10\,$ms, the toroidal field is slightly predominant over the other components, although still highly turbulent.

\begin{figure}
	\centering
	\includegraphics[width=0.4\textwidth]{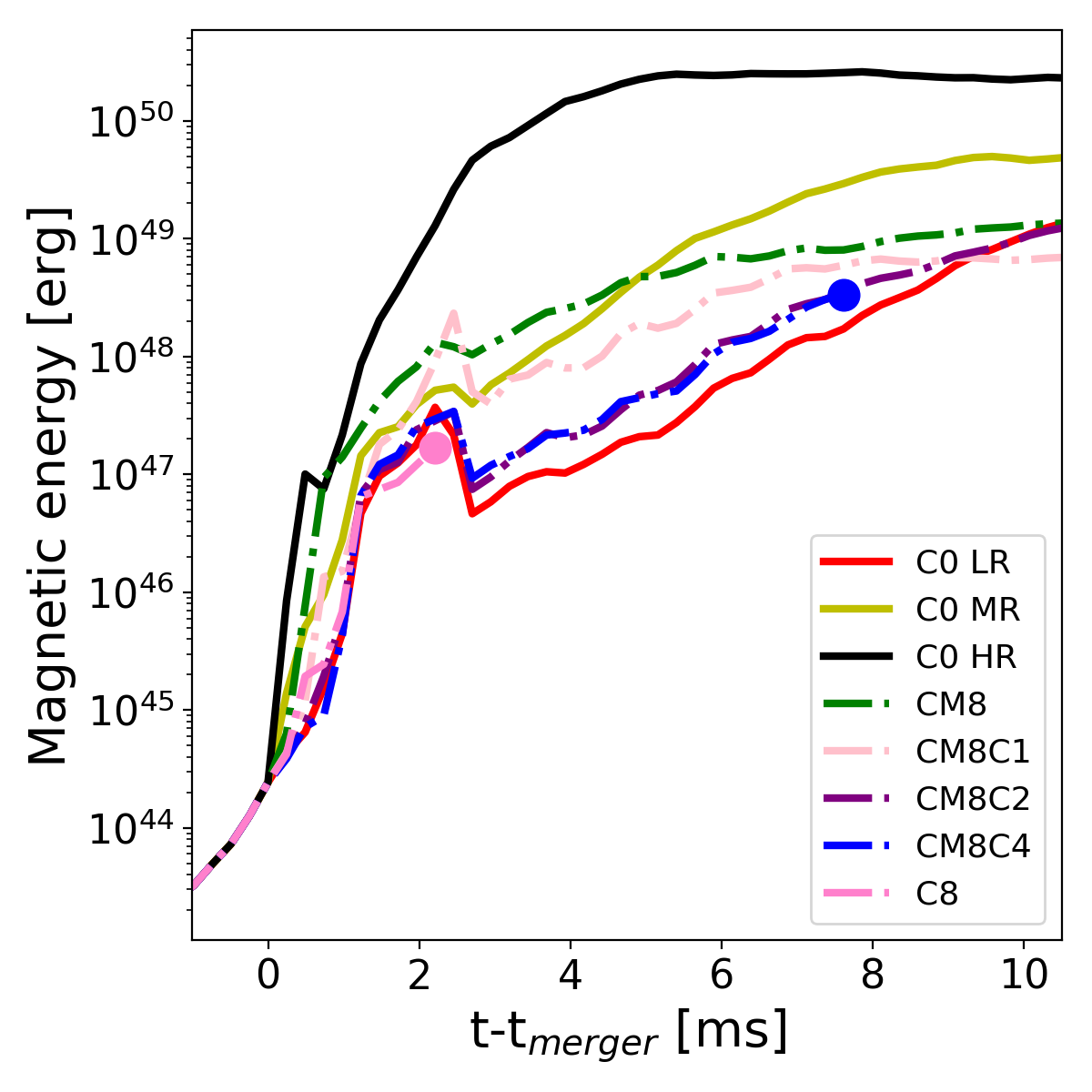}\\
	\includegraphics[width=0.4\textwidth]{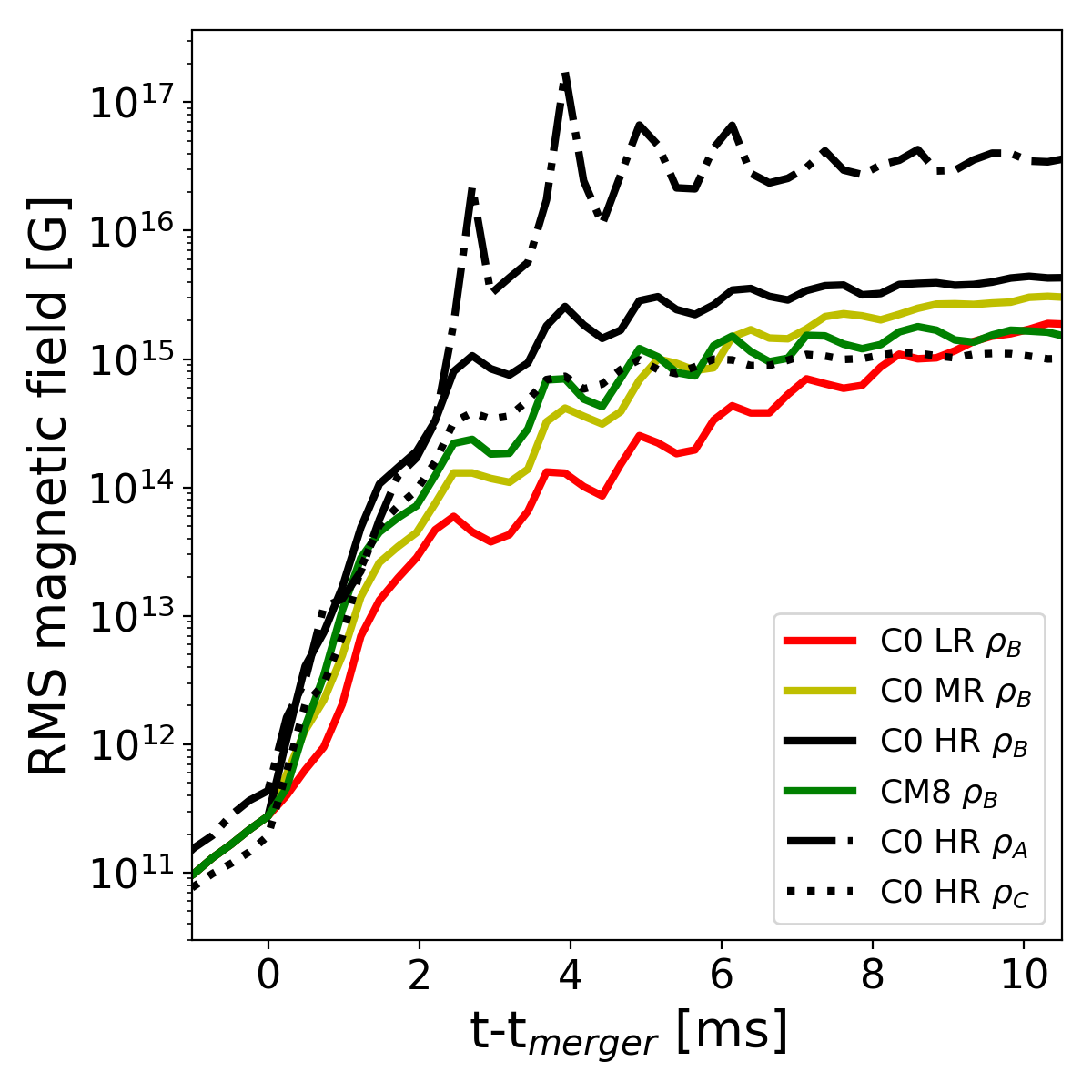}
	\caption{{\em Comparison of iLES and LES}. (Top) Integrated magnetic energy as a function of time since the merger of the BNS system. The circles indicate the collapse of the remnant, forming a black hole.
	(Bottom) R.m.s. value of the magnetic field for the iLES with different resolutions and for the most favourable LES case {\tt CM8}. The r.m.s. magnetic field of the high-resolution case, {\tt C0 HR}, is calculated in different regions with $\rho > \rho_X~\rm{g~cm^{-3}}$ , being $\rho_A= 6 \times 10^{9}$, $\rho_B= 6 \times 10^{10}$ (value used for the top panel) and $\rho_C= 6 \times 10^{11}$.}
	\label{fig:integrated_rho_magnetic}
\end{figure}

More quantitatively, Fig.~\ref{fig:spectra_tor_pol_iLES} shows the toroidal magnetic spectra, defined as the azimuthal component of the field (in simple words, the direction identified by the remnant's bulk rotation), and the poloidal one, defined by the remaining directions (the naming of such decomposition is strictly correct only in axial symmetry, but we adopt it here for simplicity). At $t=5\,$ms these two components are very similar for all resolutions. However, at later times, the toroidal field starts to grow more in the {\tt HR} case, in agreement with the above-mentioned intuition from Fig.~\ref{fig:slice_magnetic_tor_L7}.
The isotropy of the KHI-induced turbulence at early stages and the hints for a gradual ordering in the azimuthal direction at $t=10\,$ms is consistent with previous results by \cite{kiuchi18}, who showed how magnetic winding start to play an important role at this stage.

Fig.~\ref{fig:integrated_rho_magnetic} further summarizes the main results. The top panel shows the amplification of the total integrated magnetic energy for all simulations specified on Table \ref{tab:models}. Let us first focus on the iLESs (solid lines). The {\tt HR} case grows much faster, since smaller scales are excited by the KHI. This qualitatively agrees with the exponential growth rate $\propto 1/\Delta$ predicted by the KHI analytical theory \cite{miura82} and seen in previous GRMHD results \cite{kiuchi15}.
At $t=5\,$ms after merger, only the {\tt C0 HR} run has reached magnetic saturation, approximately at $2\times 10^{50}\,$erg. At this time, the magnetic energy of the three iLESs are separated by more than an order of magnitude among them.
Later on, at $t=10\,$ms, the difference on the magnetic energy between the three resolutions
is reduced almost by half, suggesting similar saturation levels of the magnetic field at late times.

The bottom panel displays the root mean square (r.m.s.)  magnetic field for the three resolution iLES and for the LES simulation with the optimal parameters, {\tt CM8}, that will be discussed in the next subsection.
The r.m.s. is computed on regions with $\rho > \rho_X~\rm{g~cm^{-3}}$, being $\rho_A= 6 \times 10^{9}$, $\rho_B= 6 \times 10^{10}$ and $\rho_C= 6 \times 10^{11}$, and taking $\rho_B$ for the magnetic energy in the top panel. The run {\tt C0 HR} shows mean values of $10^{15}$ G when considering only the most dense part of the star (i.e., $\rho > \rho_C$), but increases to $10^{16}$ G  when also the outer envelope is taken into account (i.e., $\rho > \rho_A$).

\subsection{LESs with gradient SGS model}

\begin{figure*}
	\centering
	\includegraphics[width=0.36\textwidth]{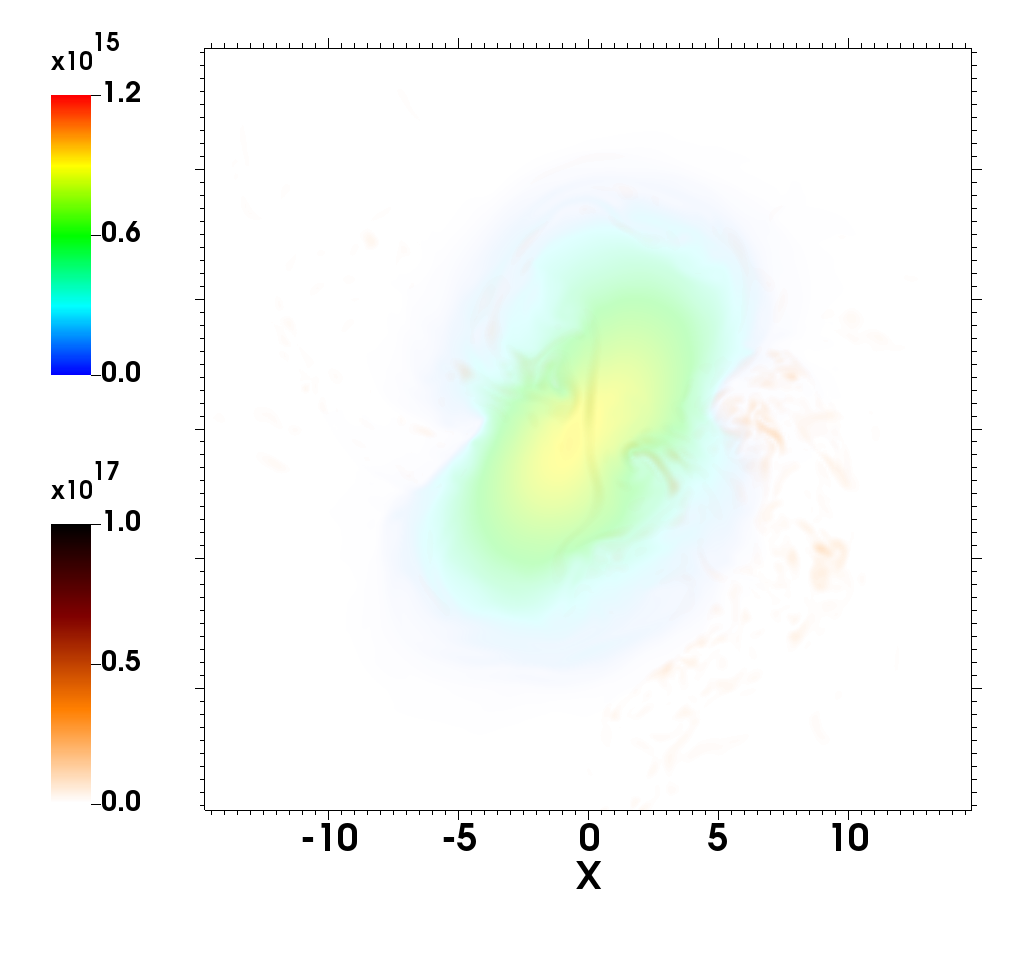}
	\includegraphics[width=0.3\textwidth,trim=170 0 0 0,clip]{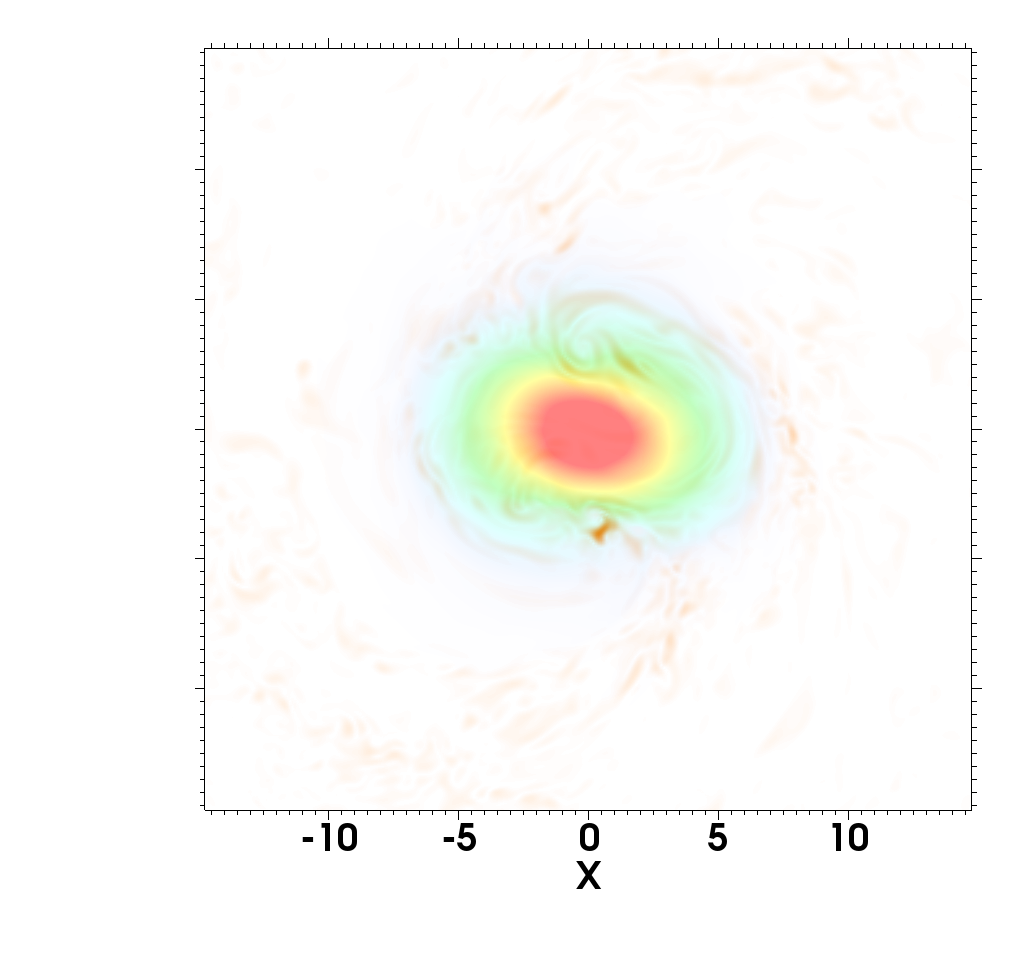}
	\includegraphics[width=0.3\textwidth,trim=170 0 0 0,clip]{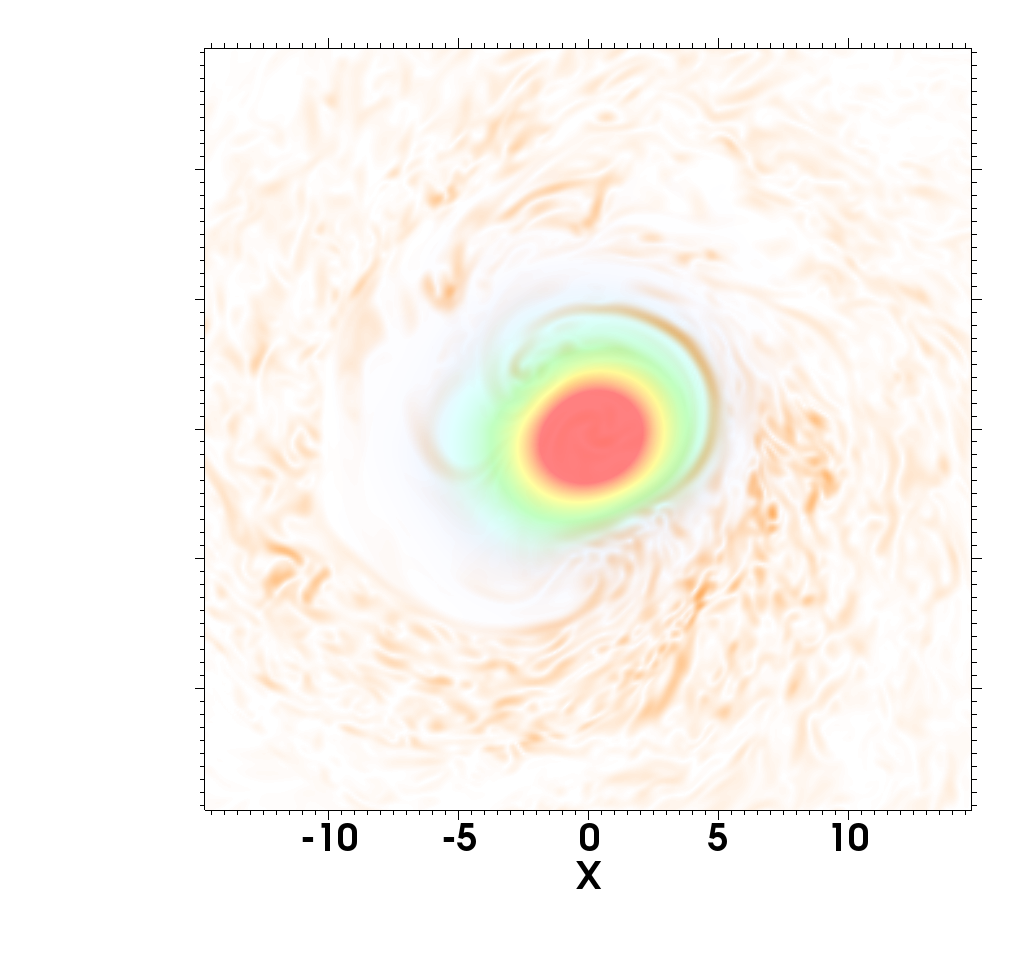} \\
	\includegraphics[width=0.36\textwidth]{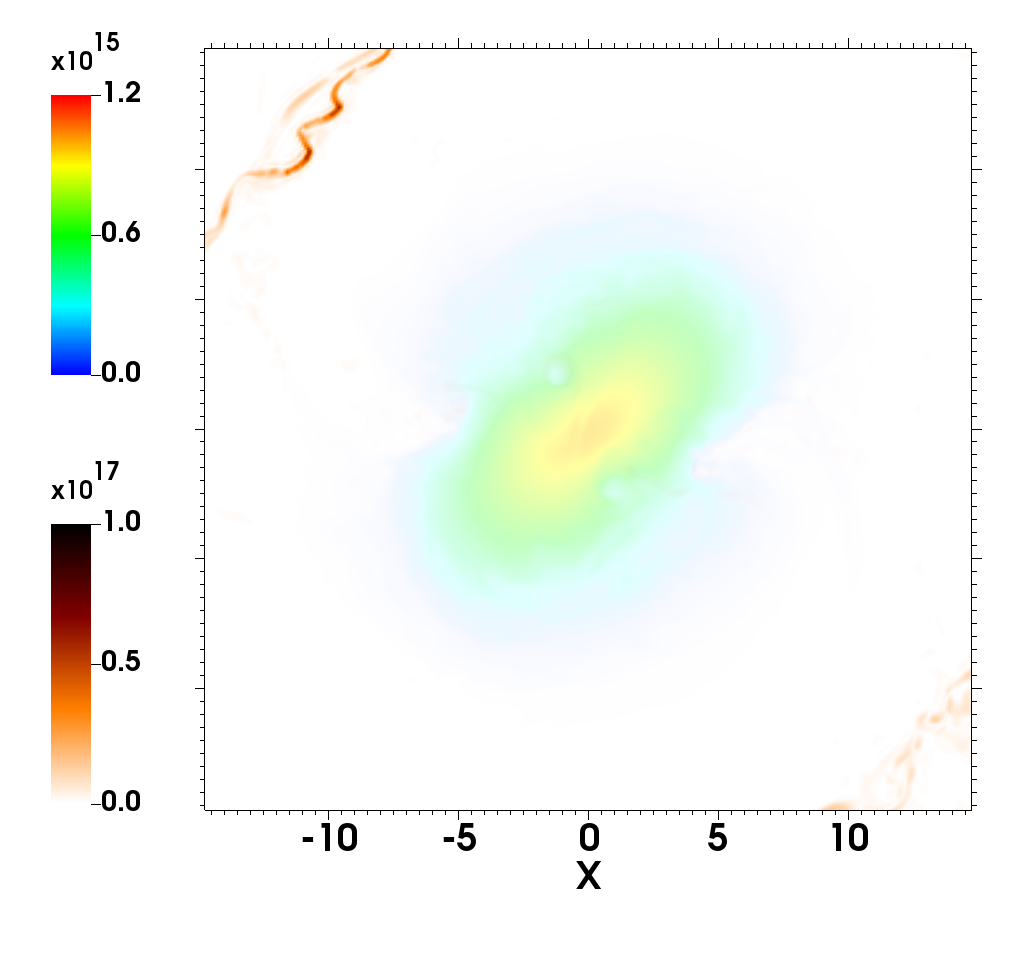}
	\includegraphics[width=0.3\textwidth,trim=170 0 0 0,clip]{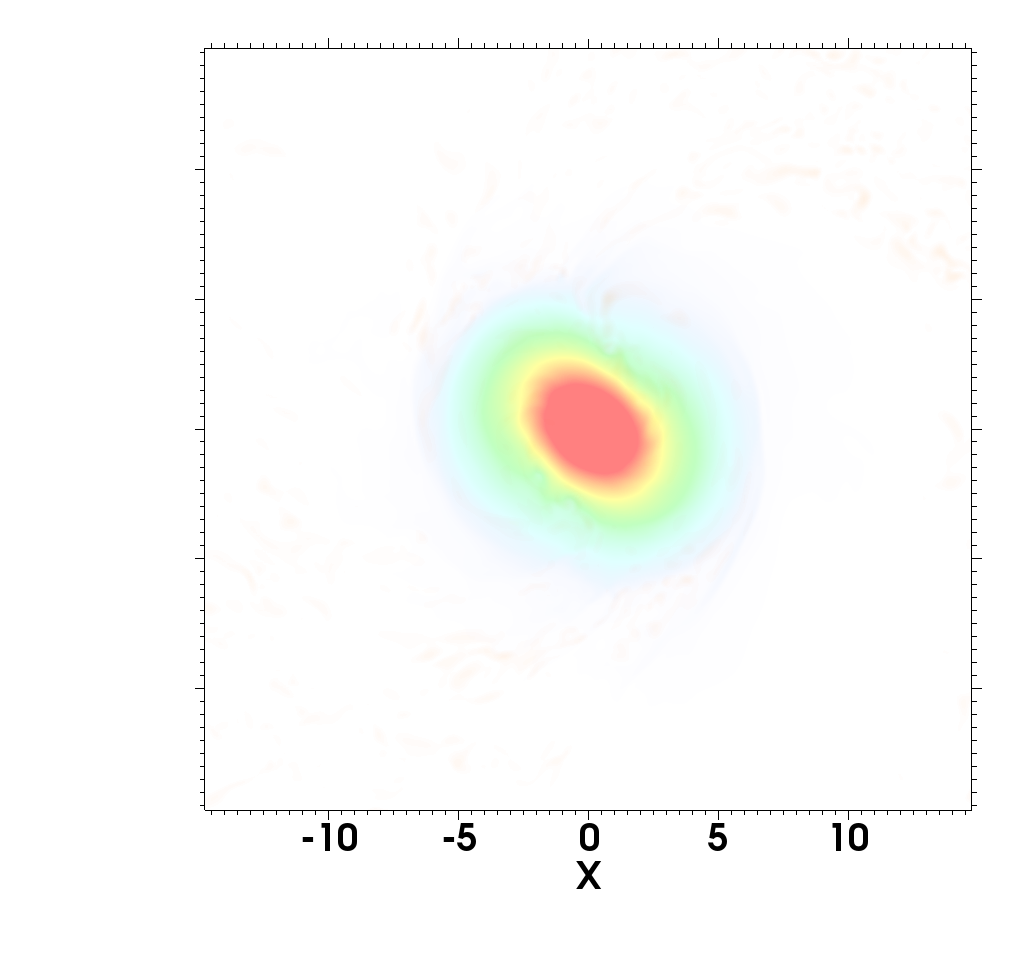}
	\includegraphics[width=0.3\textwidth,trim=170 0 0 0,clip]{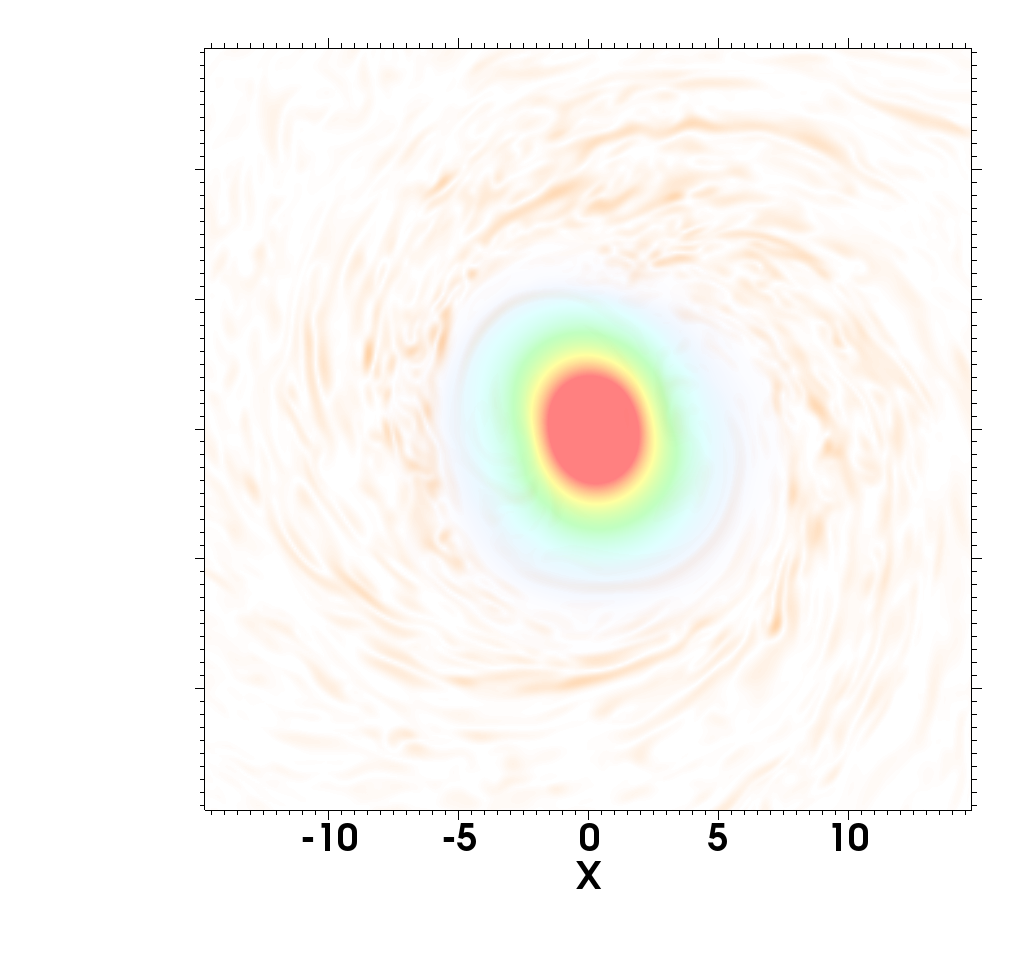}	
	\caption{{\em LES with different values of $C_i$.} Magnetic field of {\tt CM8} (top) and {\tt CM8C1} (bottom) at $t=2.5\ ms$ (left), $t=5\ ms$ (centre) and $t=10\ ms$ (right) after the merger.}
	\label{fig:slice_magnetic_CM8}
\end{figure*}

Let us now turn to the effects of including the SGS model (LESs), and continue our analysis of Fig.~\ref{fig:integrated_rho_magnetic}. We shall stress that the aim of the SGS model, applied on the {\tt LR} setup (at this particular stage of the merger evolution), is to reproduce the magnetic field amplification observed on the higher resolution simulations {\tt C0 MR/HR}.

All the {\tt LR} LESs with ${\cal C_M}=8$ show an enhanced growth in magnetic energy compared to the {\tt C0 LR}. However, we find that increasing the value of $\mathcal{C_T}$ tends to reduce the observed magnetic field amplification, presumably due to an additional effective viscosity included in the momentum equation. 
The magnetic growth is thus more prominent in the {\tt CM8} (i.e., when SGS terms are included only in the induction equation), with its integrated energy being as high as the {\tt C0 MR} run at $t=5\,$ms. The r.m.s. magnetic field for the {\tt CM8} case is comparable to {\tt C0 MR} for most of the times, and significantly larger than the {\tt C0 LR} at $t=5$ ms, although all the simulations seems to reach comparable values at $t=10$ ms, as it occurred with the total integrated magnetic energy.
Notice that the effect of the gradient SGS model is most pronounced on the initial startup stage of the KHI, whereas the magnetic growth-rates on the kinematic phase does not seem to deviate much from the {\tt C0 LR}, at it can be observed at $t=10\,$ms.

In the simulations {\tt CM8}, {\tt CM8C1} and {\tt CM8C2}, the remnant approaches a quasi-stationary stage at late times. Instead, {\tt CM8C4} and {\tt C8} show a different qualitative behavior and collapse to a black hole only after few milliseconds after the merger. This dependence on to the parameters of the SGS model is analogous to the sensitivity of the collapse time for short-lived HMNSs with numerical resolution, which has been observed previously both in HD \cite{paschalidis2015one, east2016relativistic} and MHD \cite{giacomazzo11} simulations. Notice also that increasing these parameters $\mathcal{C_T}$ and $\mathcal{C_N}$ above 2 reduce the growth of the magnetic field energy.

Analyzing more in depth what the SGS model actually does, Fig.~\ref{fig:slice_magnetic_CM8} displays the density and the magnetic field magnitude in the orbital plane $z=0$, for the {\tt CM8} (top) and {\tt CM8C1} (bottom) cases, at $t=\{2.5,5,10\}$ ms (from left to right). In both cases, the {\tt LR} by construction does not allow the formation of very fine structures like the ones of {\tt HR} (see bottom panels of Fig.~\ref{fig:slice_magnetic_iLES}). However, despite the lack of resolution, the SGS model is able to provide a growth of magnetic field up to local maximum values of $\sim 10^{16}\,$G at $t=5\,$ms, earlier than in the {\tt C0 LR}. Also for these cases, filamentary structures start to appear at about $t=10\,$ms.

\begin{figure*}
	\centering
	\includegraphics[width=0.35\textwidth]{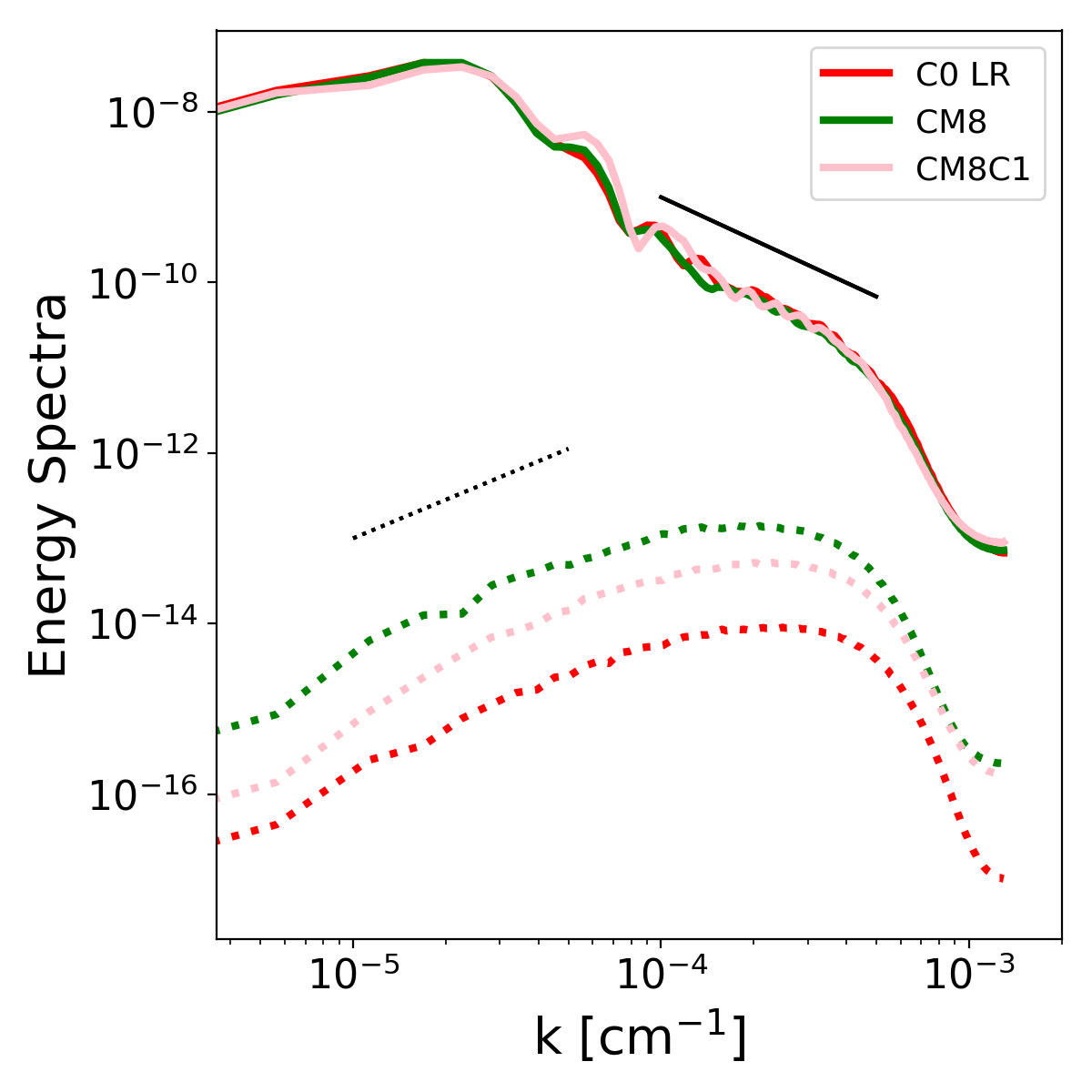}
	\includegraphics[width=0.35\textwidth]{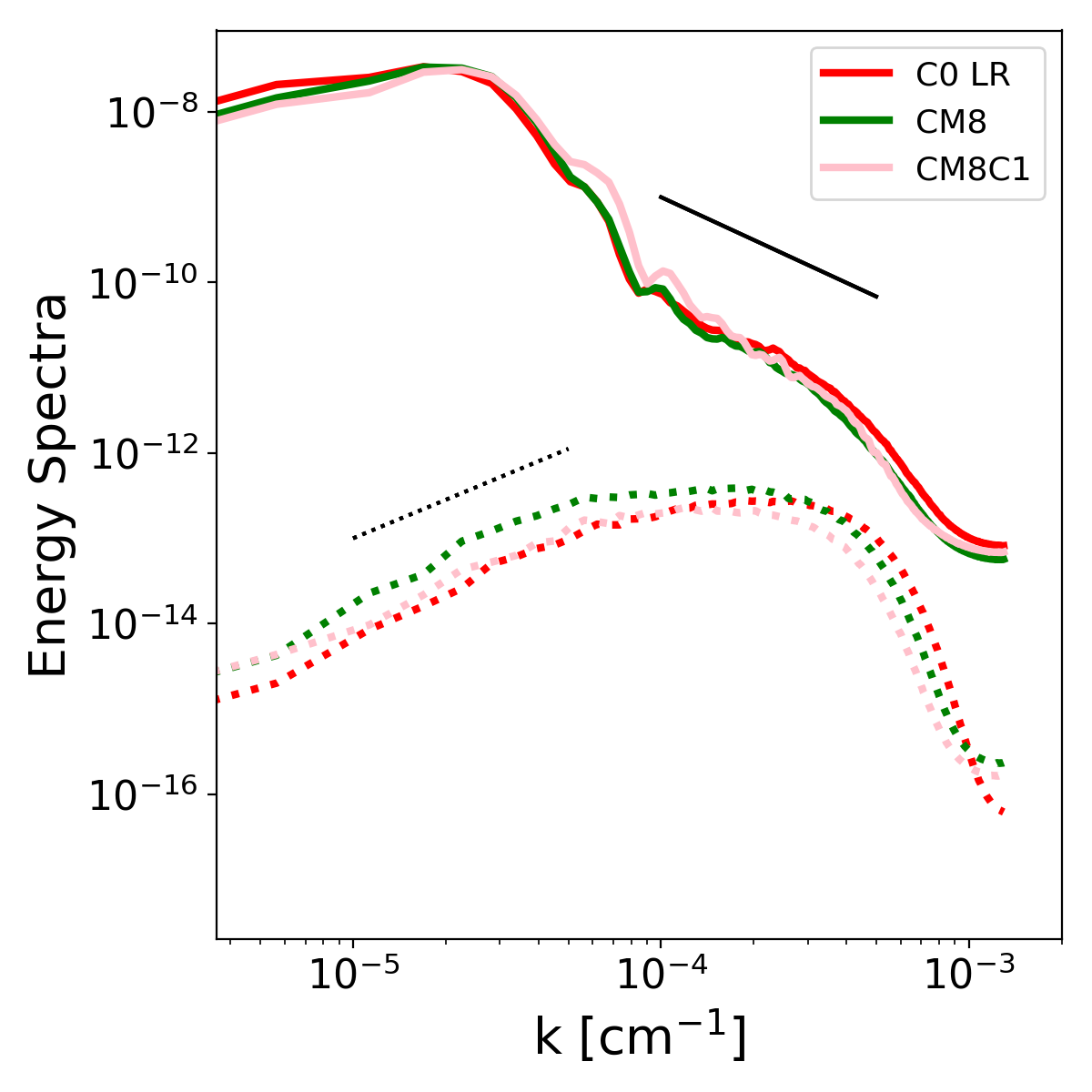}
	\caption{{\em Spectra for LESs.} Magnetic and kinetic spectra of {\tt C0 LR}, {\tt CM8} and {\tt CM8C1} at $t=5\ ms$ (left) and $t=10\ ms$. Units and black slopes as in Fig.~\ref{fig:spectra_iLES}.
	}
	\label{fig:spectra_LES}
\end{figure*}

A comparison among the spectra is shown in Fig.~\ref{fig:spectra_LES}, for {\tt CM8}, {\tt CM8C1} and {\tt C0 LR}, at $t=\{5,10\}$ ms after merger. Overall, these profiles are similar to those of iLESs, with the main difference given by their integrated values (i.e., the total magnetic energy). This again shows that at $t=5\,$ms {\tt CM8} is the most amplified one among the {\tt LR} cases, between two and three orders of magnitude higher than the others for all wavenumbers (except the very high ones, which are dominated by numerical dissipation). The {\tt CM8C1} case exhibit a moderate growth of the magnetic energy spectra with respect to the {\tt C0 LR} run, but considerably smaller than {\tt CM8}.
At $t=10\,$ms, 
the spectral distribution for these three cases is quite similar, and very close to equipartition at large wavenumbers. This is again consistent with Fig.~\ref{fig:integrated_rho_magnetic}, where these low-resolution simulations reach nearly the same magnetic energy values at late times. This behaviour on the magnetic energy spectra of LES was also found in our bounding-box simulations \cite{vigano19b,carrasco19,vigano20}.

In summary, the {\tt LR} LES that have a closer resemblance to the higher-resolution iLES (i.e., {\tt C0 MR/HR}), at least at these early times, is {\tt CM8}.

\section{Conclusions}\label{sec:conclusions}

In this article, we showed the first results from LESs of BNS with the extended gradient model, already presented for non-relativistic \cite{vigano19b}, special \cite{carrasco19} and general relativistic \cite{vigano20} MHD box simulations of the KHI; here it has been implemented in a full GRMHD code in order to study the BNS merger scenario. Moreover, our code implements overall fourth-order accurate numerical schemes, while most existing GRMHD simulations rely on second-order accurate approaches (see advantages in the use of fourth-order schemes in \cite{most19}).

We have focused on the the magnetic field amplification within the first $\sim10\,$ ms after the BNS merger.
And analyzed the role of numerical resolution in capturing these MHD turbulent-dynamo effects.
With our best-resolved run reaching a grid-spacing of $\Delta \sim 37\,$m (in the finest level) and relying on the use of high-order numerical methods, for our highest resolution simulation we were able to demonstrate an amplification of r.m.s. values between $10^{15}$ G, in the densest regions of the remnant, and $10^{16}$ G, when the less dense outer envelope is also considered.
 
We have tested the gradient SGS model, by studying different values of its  pre-coefficients $\mathcal{C}_i$, one in each equation. We show that the SGS terms on the induction equation acting in a moderate resolution (with $\Delta \sim 147\,$m) are able to mimic at least the magnetic growth of a better-resolved simulation (with $\Delta \sim 74\,$m). 

In our previous works \cite{vigano19b,vigano20} we observed that the best results, with the gradient SGS model and our numerical schemes, were achieved by setting the constants $\mathcal{C}_i$ approximately up to one order of magnitude larger than their theoretical values $\mathcal{C}_i=1$. We concluded this was due to the intrinsic dissipation of our numerical scheme. 
However, in the present context, we found that if we set all $\mathcal{C}_i=8$ there is an excessive dissipation in the momentum equation which prevents a rapid growth of the magnetic field during the turbulent regime, whereas it accelerates the collapse to a black hole of the remnant. Taking this into account, we have found that the best calibration (in terms of reproducing the higher-resolution magnetic field amplification) consists in rather high values ${\cal C_M}=8$ but ${\cal C_T}={\cal C_N}\sim 0$--$2$. 
The reason for this remain to be clarified, and probably lies in the presence of two scales in the system. First, a fairly well resolved hydrodynamical one which includes the differential rotation and convection within the remnant. Second, a much smaller MHD scale, involving the turbulent dynamics originated by the KHI, which is still far from
being resolved. As a consequence, only the coefficients for the magnetic field evolution needs to be artificially enhanced from their theoretical values, in order to maximize the effect of the SGS model to approach the results obtained with higher resolution simulations.
Introducing too large contributions of the SGS model in the momentum equation leads to a higher effective viscosity which finally hampers the turbulence, partially suppressing the magnetic field amplification. In order to understand better the details and to disentangle the numerical and physical reasons for this, it would be helpful to implement and test our  SGS model in other codes, for different numerical methods and/or scenarios.

Regardless on the details of the SGS modeling, we have shown how the energy and magnetic spectra follow, respectively, the expected Kolmogorov ($k^{-5/3}$) and Kazantsev ($k^{3/2}$) slopes, as in our iLES simulations. The magnetic spectra have a peak at small k, very different from a large-scale ordered field. Therefore, we warn against the widespread argument that an initially strong large-scale
magnetic field can compensate the lack of ability to follow the KHI growth: the latter is intrinsically turbulent and can easily provide local maximum values exceeding $10^{17}$ G, but contained in very small structures.

This is also consistent with the fact that at early times where the kinetic dynamics deriving from the collision of the two cores is still dominating. The KHI triggers a quite isotropic turbulence, which destroys any large-scale weak field. Only at later times, the dominating differential rotation should provide (via winding and MRI) the necessary energy injection at large scales which could partially order such strong but finely structured magnetic field. This can happen via inverse cascade and isotropy breaking favouring in particular the stretching of magnetic field lines in the azimuthal direction.
We hope that our approach, which is applicable to any GRMHD problem, can be used in the near future to explore the post-merger phase dynamics.
Given its potential in capturing the turbulent-dynamo effect at a much lower computational costs, it could be useful to further assess the production of large-scale fields that is required for the jet formation associated to the SGRB.

\subsection*{Acknowledgments} 
We thank Wolfgang Kastaun for his help and assistance on the implementation of their recovery procedure~\cite{kastaun20} into our evolution code.
We acknowledge support from the Spanish Ministry of Economy and Competitiveness grants AYA2016-80289-P and PID2019-110301GB-I00 (AEI/FEDER, UE). DV is supported by the ERC Consolidator Grant "MAGNESIA" (nr. 817661) and the Spanish grant SGR~2017-1383. We thank the technical support provided by Barcelona Supercomputing Center, where most computational resources used for this work have been granted to our project LESBNS by the $20^{th}$ PRACE regular call (Proposal 2019215177, P.I. CP and DV). 

\appendix

\section{Explicit form of the gradient sub-grid tensors}\label{app:sgs_forms}

The explicit expression of the $H$ tensors appearing in the SGS gradient terms were first obtained in Ref.~\cite{carrasco19}, and then extended to GR in Ref.~\cite{vigano20}. They can be written as
\begin{widetext}
	\begin{eqnarray}
	 {\Psi}_{v}^k &=& \frac{2}{ {\Theta}} \left\lbrace \nabla ( {v}\cdot  {B}) \cdot \nabla  {B}^k  - \nabla  {\Theta} \cdot \nabla  {v}^k   
	+ \frac{ {B}^k}{ {\mathcal{E}}} \left[   {\Theta} \nabla  {B}^j \cdot \nabla  {v}_j +  {B}_j \nabla  {B}^j \cdot \nabla ( {v}\cdot  {B}) -  {B}^j \nabla  {v}_j \cdot \nabla  {\Theta} \right]  \label{hTauv}  \right\rbrace ~, \nonumber  \\
	 {\Psi}^{ki}_{M} &=& \frac{4}{ {\Theta}} \left[   {\Theta} \, \nabla  {B}^{[i} \cdot \nabla  {v}^{k]} +   {B}^{[i} \nabla  {B}^{k]} \cdot \nabla ( {v}\cdot  {B}) -  {B}^{[i} \nabla  {v}^{k]} \cdot \nabla  {\Theta} \right] ~,
	\nonumber \\
	 {\Psi}_{\Theta} &=&  \frac{ {\Theta}}{ {\Theta} - {E}^2} \left\lbrace \nabla  {B}_{j} \cdot \nabla  {B}^{j} - \nabla  {E}_{j} \cdot \nabla  {E}^{j} -  {B}_{[i} {v}_{k]} \,  {\Psi}^{ki}_{M} \right\rbrace ~~,~~
	 {\Psi}_{A}  =  {W}^2 \left(  {p} \, \frac{d {p}}{d {\epsilon}} +  {\rho}^2 \, \frac{d {p}}{d {\rho}} \right) ~, \nonumber \\
	H_{\rm p} &=&  \frac{ {\mathcal{E}} \,  {W}^2 ({ {\Theta}-  {E}^2 })}{( {\rho} \,  {\mathcal{E}} -  {\Psi}_{A})( {\Theta} -  {E}^2 )  {W}^2 +  {\Psi}_{A} \,  {\Theta}} \left\lbrace  {\rho} \left( \nabla \frac{d {p}}{d {\rho}} \cdot \nabla  {\rho} + \nabla \frac{d {p}}{d {\epsilon}} \cdot \nabla  {\epsilon} \right)  - 2 \frac{d {p}}{d {\epsilon}} \, \nabla  {\rho} \cdot \nabla  {\epsilon}  \right. \nonumber\\
	&-&  \left.  \left( {\mathcal{E}} \frac{d {p}}{d {\epsilon}} -  {\Psi}_{A}\right) \left[ \frac{ {W}^2}{4} \nabla  {W}^{-2} \cdot \nabla  {W}^{-2} + \nabla  {W}^{-2} \cdot \nabla (\ln  {\rho}) \right]  -  \frac{2}{ {W}^2}\frac{d {p}}{d {\epsilon}} \left[   \nabla  {B}_j \cdot \nabla  {B}^j -   {W}^{4} \nabla  {W}^{-2} \cdot \nabla  {h} \right]  \right. \label{tau_p} \\
	&-&  \left.  \left( {\mathcal{E}} \frac{d {p}}{d {\epsilon}} +  {\Psi}_{A}\right) \left[  {v}_j  {\Psi}_{v}^j +  \nabla  {v}_{j} \cdot \nabla  {v}^{j} +  {W}^2 \, \nabla  {W}^{-2} \cdot \nabla  {W}^{-2} \right]  +   \frac{ {\Psi}_{\Theta}}{ {\mathcal{E}}  {\Theta}} \left[ \left( {\mathcal{E}} \frac{d {p}}{d {\epsilon}} +  {\Psi}_{A}\right)( {\Theta}-  {E}^2 ) - \frac{ {\Psi}_{A} \,  {\Theta}}{ {W}^2}   \right]   \right\rbrace ~, \nonumber \\
	H_{\Theta} &=&  {\Psi}_{\Theta} + \frac{ {\Theta}}{ {\Theta} - {E}^2} H_p ~, \label{tau_Theta} \\ 
	H_{v}^k &:=&  {\Psi}_{v}^k - \left(  {v}^k + \frac{ {v}\cdot  {B}}{ {\mathcal{E}}}  {B}^k \right)  \frac{H_{\Theta}}{ {\Theta}} ~, \label{Tauv} \\
	H^{k}_{N} &=&  2 \, \nabla  {D} \cdot \nabla  {v}^k +  {D} \, H^{k}_v ~~,~~
	H^{ki}_{M} =  2  {B}^{[i} H_{v}^{k]} + 4 \, \nabla  {B}^{[i} \cdot \nabla  {v}^{k]} ~~\rightarrow~~ 
	H_{E}^i = \frac{1}{2} \epsilon^{i}_{\phantom ijk } H_{M}^{jk} ~,
	\label{HNME} \\
	H^{ki}_{T} &=& 2 \left[ \nabla  {\mathcal{E}} \cdot \nabla ( {v}^k  {v}^i ) +  {\mathcal{E}} \left(  {v}^{(k} H_{v}^{i)} +  \nabla  {v}^{k} \cdot \nabla  {v}^{i} \right)  +   {v}^k  {v}^i H_{p} \right] 
	- 2\left[ \nabla  {B}^{k} \cdot \nabla  {B}^{i} + \nabla  {E}^{k} \cdot \nabla  {E}^{i} +  {E}^{(k} H_{E}^{i)}   \right]  \nonumber \\
	&+& (\gamma^{ki} -  {v}^k  {v}^i)  \left[ H_p + \nabla  {B}_{j} \cdot \nabla  {B}^{j} + \nabla  {E}_{j} \cdot \nabla  {E}^{j} +  {E}_{j} H_{E}^{j} \right]~, \label{HT}     
	\end{eqnarray}
\end{widetext}
where $h= \rho (1 + \epsilon) + p$ is the enthalpy, $\mathcal{E} = h W^2$, 
$\Theta = \mathcal{E} + B^2$, and the two gradients $\nabla$ (on each term) symbolize spatial partial derivatives $\partial_i$ (and $\partial_j$), with ``$\cdot$'' indicating contraction among them with the spatial metric $\gamma^{ij}$.

\section{Spectra calculation}\label{app:spectra}

It is illustrative to compute the radially-averaged spectrum of the kinetic and magnetic energy \cite{durran17,vigano19b}. For a given field $f$ defined in a periodic box of side $L$, we use common {\tt python} functions to calculate its discrete fast Fourier transform $\hat{f}(\vec{k}) = \Sigma_{\vec{x}} f(\vec{x}) e^{-i \vec{k}\cdot\vec{x}}$, where the sum is performed over the $N^3$ points equally spaced in each direction, with $k_j = n~\Delta k$, where $\Delta k = \frac{2\pi}{L}$ and $n\in [0,N/2]$ is an integer. Then, we calculate the solid-angle-averaged values $4\pi < k^2 |f|^2 >_{k}$ over the radial bins in the Fourier space, centered at $k=\{n~\Delta k\}$, which represent the power density per unit of angular wavenumber. This defines the kinetic and magnetic spectra,
\begin{eqnarray}
&& \mathcal{E}_k(k) = \frac{L^3 4\pi}{(2\pi)^3N^6} <k^2|\widehat{\sqrt{\rho}\vec{v}}|^2(\vec{k})>_{k}~, \nonumber\\
&& \mathcal{E}_m(k) = \frac{L^3 4\pi }{(2\pi)^3N^6}<k^2|\hat{\vec{B}}|^2(\vec{k})>_{k}~, ~\label{eq:spectra}
\end{eqnarray}
that we define for simplicity as in the non-relativistic case.

The calculation of the spectra is done by choosing the same 70-km-wide cube of the fifth FMR level, which encloses the remnant (i.e., almost the totality of the kinetic and magnetic energy of the system). Within this domain, the information analyzed is the one of the finest grid available. Since the domains of the AMR levels in the {\tt MR} and {\tt HR} cases are smaller than such domain over which spectra are calculated, we interpolate the values of the fields from the coarser levels, filling a regular mesh with the same grid spacing as the highest level present in that simulation. By construction, such interpolation have effects on the spectra limited to the smallest scales (highest $k$), which are not resolved outside the AMR levels.
In addition, and in order to reduce the contamination from the rarefied  atmosphere, the spectra is computed by considering only the velocity and magnetic fields in the regions with density larger than $6 \times 10^{9}$ g cm$^{-3}$, setting them to zero otherwise.

\bibliographystyle{unsrt}
\bibliography{turbulence}

\begin{thebibliography}{10}

\bibitem{abbott17a}
B.~P. {Abbott}, R.~{Abbott}, T.~D. {Abbott}, F.~{Acernese}, K.~{Ackley},
  C.~{Adams}, T.~{Adams}, P.~{Addesso}, R.~X. {Adhikari}, V.~B. {Adya}, and
  et~al.
\newblock {GW170817: Observation of Gravitational Waves from a Binary Neutron
  Star Inspiral}.
\newblock {\em Physical Review Letters}, 119(16):161101, October 2017.

\bibitem{abbott17b}
B.~P. {Abbott}, R.~{Abbott}, T.~D. {Abbott}, F.~{Acernese}, K.~{Ackley},
  C.~{Adams}, T.~{Adams}, P.~{Addesso}, R.~X. {Adhikari}, V.~B. {Adya}, and
  et~al.
\newblock {Multi-messenger Observations of a Binary Neutron Star Merger}.
\newblock {\em \apjl}, 848:L12, October 2017.

\bibitem{goldstein2017}
A~Goldstein, P~Veres, E~Burns, MS~Briggs, R~Hamburg, D~Kocevski,
  CA~Wilson-Hodge, RD~Preece, S~Poolakkil, OJ~Roberts, et~al.
\newblock An ordinary short gamma-ray burst with extraordinary implications:
  Fermi-gbm detection of grb 170817a.
\newblock {\em The Astrophysical Journal Letters}, 848(2):L14, 2017.

\bibitem{savchenko2017}
V~Savchenko, C~Ferrigno, E~Kuulkers, A~Bazzano, E~Bozzo, S~Brandt, J~Chenevez,
  TJ-L Courvoisier, R~Diehl, A~Domingo, et~al.
\newblock Integral detection of the first prompt gamma-ray signal coincident
  with the gravitational-wave event gw170817.
\newblock {\em The Astrophysical Journal Letters}, 848(2):L15, 2017.

\bibitem{abbott17d}
B.~P. {Abbott}, R.~{Abbott}, T.~D. {Abbott}, F.~{Acernese}, K.~{Ackley}, and
  et~al.
\newblock {Gravitational Waves and Gamma-Rays from a Binary Neutron Star
  Merger: GW170817 and GRB 170817A}.
\newblock {\em \apjl}, 848(2):L13, October 2017.

\bibitem{abbott17c}
B.~P. {Abbott}, R.~{Abbott}, T.~D. {Abbott}, F.~{Acernese}, K.~{Ackley},
  C.~{Adams}, T.~{Adams}, P.~{Addesso}, R.~X. {Adhikari}, V.~B. {Adya}, and
  et~al.
\newblock {Estimating the Contribution of Dynamical Ejecta in the Kilonova
  Associated with GW170817}.
\newblock {\em \apjl}, 850:L39, December 2017.

\bibitem{metzger17}
B.~D. {Metzger}.
\newblock {Kilonovae}.
\newblock {\em Living Reviews in Relativity}, 20:3, May 2017.

\bibitem{davanzo2018}
P~D’avanzo, S~Campana, OS~Salafia, G~Ghirlanda, G~Ghisellini, A~Melandri,
  MG~Bernardini, M~Branchesi, E~Chassande-Mottin, S~Covino, et~al.
\newblock The evolution of the x-ray afterglow emission of gw 170817/grb
  170817a in xmm-newton observations.
\newblock {\em Astronomy \& Astrophysics}, 613:L1, 2018.

\bibitem{fong2019}
Wen-fai Fong, PK~Blanchard, KD~Alexander, J~Strader, Raffaella Margutti,
  A~Hajela, VA~Villar, Y~Wu, CS~Ye, E~Berger, et~al.
\newblock The optical afterglow of gw170817: An off-axis structured jet and
  deep constraints on a globular cluster origin.
\newblock {\em The Astrophysical Journal Letters}, 883(1):L1, 2019.

\bibitem{dobie2018}
Dougal Dobie, David~L Kaplan, Tara Murphy, Emil Lenc, Kunal~P Mooley, Christene
  Lynch, Alessandra Corsi, Dale Frail, Mansi Kasliwal, and Gregg Hallinan.
\newblock A turnover in the radio light curve of gw170817.
\newblock {\em The Astrophysical Journal Letters}, 858(2):L15, 2018.

\bibitem{mooley2018}
KP~Mooley, DA~Frail, D~Dobie, E~Lenc, A~Corsi, K~De, AJ~Nayana, S~Makhathini,
  I~Heywood, T~Murphy, et~al.
\newblock A strong jet signature in the late-time light curve of gw170817.
\newblock {\em The Astrophysical Journal Letters}, 868(1):L11, 2018.

\bibitem{margalit17}
B.~{Margalit} and B.~D. {Metzger}.
\newblock {Constraining the Maximum Mass of Neutron Stars from Multi-messenger
  Observations of GW170817}.
\newblock {\em \apjl}, 850:L19, December 2017.

\bibitem{shibata2017modeling}
Masaru Shibata, Sho Fujibayashi, Kenta Hotokezaka, Kenta Kiuchi, Koutarou
  Kyutoku, Yuichiro Sekiguchi, and Masaomi Tanaka.
\newblock Modeling gw170817 based on numerical relativity and its implications.
\newblock {\em Physical Review D}, 96(12):123012, 2017.

\bibitem{abbott2018}
Benjamin~P Abbott, Richard Abbott, TD~Abbott, F~Acernese, K~Ackley, C~Adams,
  T~Adams, P~Addesso, Rana~X Adhikari, Vaishali~B Adya, et~al.
\newblock Gw170817: Measurements of neutron star radii and equation of state.
\newblock {\em Physical review letters}, 121(16):161101, 2018.

\bibitem{palenzuela2013electromagnetic}
Carlos Palenzuela, Luis Lehner, Marcelo Ponce, Steven~L Liebling, Matthew
  Anderson, David Neilsen, and Patrick Motl.
\newblock Electromagnetic and gravitational outputs from binary-neutron-star
  coalescence.
\newblock {\em Physical review letters}, 111(6):061105, 2013.

\bibitem{kiuchi14}
K.~{Kiuchi}, K.~{Kyutoku}, Y.~{Sekiguchi}, M.~{Shibata}, and T.~{Wada}.
\newblock {High resolution numerical relativity simulations for the merger of
  binary magnetized neutron stars}.
\newblock {\em \prd}, 90(4):041502, August 2014.

\bibitem{neilsen2014magnetized}
David Neilsen, Steven~L Liebling, Matthew Anderson, Luis Lehner, Evan
  O’Connor, and Carlos Palenzuela.
\newblock Magnetized neutron stars with realistic equations of state and
  neutrino cooling.
\newblock {\em Physical Review D}, 89(10):104029, 2014.

\bibitem{kiuchi15}
K.~{Kiuchi}, P.~{Cerd{\'a}-Dur{\'a}n}, K.~{Kyutoku}, Y.~{Sekiguchi}, and
  M.~{Shibata}.
\newblock {Efficient magnetic-field amplification due to the Kelvin-Helmholtz
  instability in binary neutron star mergers}.
\newblock {\em \prd}, 92(12):124034, December 2015.

\bibitem{giacomazzo15}
B.~{Giacomazzo}, J.~{Zrake}, P.~C. {Duffell}, A.~I. {MacFadyen}, and
  R.~{Perna}.
\newblock {Producing Magnetar Magnetic Fields in the Merger of Binary Neutron
  Stars}.
\newblock {\em \apj}, 809:39, August 2015.

\bibitem{palenzuela15}
C.~{Palenzuela}, S.~L. {Liebling}, D.~{Neilsen}, L.~{Lehner}, O.~L.
  {Caballero}, E.~{O'Connor}, and M.~{Anderson}.
\newblock {Effects of the microphysical equation of state in the mergers of
  magnetized neutron stars with neutrino cooling}.
\newblock {\em \prd}, 92(4):044045, August 2015.

\bibitem{ruiz16}
M.~{Ruiz}, R.~N. {Lang}, V.~{Paschalidis}, and S.~L. {Shapiro}.
\newblock {Binary Neutron Star Mergers: A Jet Engine for Short Gamma-Ray
  Bursts}.
\newblock {\em \apjl}, 824:L6, June 2016.

\bibitem{kiuchi18}
K.~{Kiuchi}, K.~{Kyutoku}, Y.~{Sekiguchi}, and M.~{Shibata}.
\newblock {Global simulations of strongly magnetized remnant massive neutron
  stars formed in binary neutron star mergers}.
\newblock {\em \prd}, 97(12):124039, June 2018.

\bibitem{ciolfi2019}
Riccardo Ciolfi, Wolfgang Kastaun, Jay~Vijay Kalinani, and Bruno Giacomazzo.
\newblock First 100 ms of a long-lived magnetized neutron star formed in a
  binary neutron star merger.
\newblock {\em Physical Review D}, 100(2):023005, 2019.

\bibitem{ciolfi2020collimated}
Riccardo Ciolfi.
\newblock Collimated outflows from long-lived binary neutron star merger
  remnants.
\newblock {\em Monthly Notices of the Royal Astronomical Society: Letters},
  495(1):L66--L70, 2020.

\bibitem{ruiz2020}
Milton Ruiz, Antonios Tsokaros, and Stuart~L Shapiro.
\newblock Magnetohydrodynamic simulations of binary neutron star mergers in
  general relativity: Effects of magnetic field orientation on jet launching.
\newblock {\em Physical Review D}, 101(6):064042, 2020.

\bibitem{mosta2020}
Philipp M{\"o}sta, David Radice, Roland Haas, Erik Schnetter, and Sebastiano
  Bernuzzi.
\newblock A magnetar engine for short grbs and kilonovae.
\newblock {\em arXiv preprint arXiv:2003.06043}, 2020.

\bibitem{ciolfi2020key}
Riccardo Ciolfi.
\newblock The key role of magnetic fields in binary neutron star mergers.
\newblock {\em arXiv preprint arXiv:2003.07572}, 2020.

\bibitem{tauris17}
T.~M. {Tauris}, M.~{Kramer}, P.~C.~C. {Freire}, N.~{Wex}, H.~T. {Janka},
  N.~{Langer}, Ph. {Podsiadlowski}, E.~{Bozzo}, S.~{Chaty}, M.~U. {Kruckow},
  E.~P.~J. {van den Heuvel}, J.~{Antoniadis}, R.~P. {Breton}, and D.~J.
  {Champion}.
\newblock {Formation of Double Neutron Star Systems}.
\newblock {\em \apj}, 846(2):170, Sep 2017.

\bibitem{bilous2019nicer}
Anna~V Bilous, Anna~L Watts, Alice~K Harding, Thomas~E Riley, Zaven
  Arzoumanian, Slavko Bogdanov, Keith~C Gendreau, Paul~S Ray, Sebastien
  Guillot, Wynn~CG Ho, et~al.
\newblock A nicer view of psr j0030+ 0451: evidence for a global-scale
  multipolar magnetic field.
\newblock {\em The Astrophysical Journal Letters}, 887(1):L23, 2019.

\bibitem{rea2010low}
N~Rea, P~Esposito, R~Turolla, GL~Israel, S~Zane, L~Stella, S~Mereghetti,
  A~Tiengo, D~G{\"o}tz, Ersin G{\"o}{\u{g}}{\"u}{\c{s}}, et~al.
\newblock A low-magnetic-field soft gamma repeater.
\newblock {\em Science}, 330(6006):944--946, 2010.

\bibitem{obergaulinger10}
M.~{Obergaulinger}, M.~A. {Aloy}, and E.~{M{\"u}ller}.
\newblock {Local simulations of the magnetized Kelvin-Helmholtz instability in
  neutron-star mergers}.
\newblock {\em A\&A}, 515:A30, June 2010.

\bibitem{zrake13b}
J.~{Zrake} and A.~I. {MacFadyen}.
\newblock {Magnetic Energy Production by Turbulence in Binary Neutron Star
  Mergers}.
\newblock {\em \apjl}, 769:L29, June 2013.

\bibitem{guilet17}
J.~{Guilet}, A.~{Bauswein}, O.~{Just}, and H.-T. {Janka}.
\newblock {Magnetorotational instability in neutron star mergers: impact of
  neutrinos}.
\newblock {\em \mnras}, 471:1879--1887, October 2017.

\bibitem{zhiyin15}
Zhiyin Yang.
\newblock Large-eddy simulation: Past, present and the future.
\newblock {\em Chinese Journal of Aeronautics}, 91, 12 2014.

\bibitem{bucciantini13}
N.~{Bucciantini} and L.~{Del Zanna}.
\newblock {A fully covariant mean-field dynamo closure for numerical 3 + 1
  resistive GRMHD}.
\newblock {\em \mnras}, 428:71--85, January 2013.

\bibitem{duez2004}
Matthew~D Duez, Yuk~Tung Liu, Stuart~L Shapiro, and Branson~C Stephens.
\newblock General relativistic hydrodynamics with viscosity: Contraction,
  catastrophic collapse, and disk formation in hypermassive neutron stars.
\newblock {\em Physical Review D}, 69(10):104030, 2004.

\bibitem{shibata2017general}
Masaru Shibata, Kenta Kiuchi, and Yu-ichiro Sekiguchi.
\newblock General relativistic viscous hydrodynamics of differentially rotating
  neutron stars.
\newblock {\em Physical Review D}, 95(8):083005, 2017.

\bibitem{radice17}
D.~{Radice}.
\newblock {General-relativistic Large-eddy Simulations of Binary Neutron Star
  Mergers}.
\newblock {\em \apjl}, 838:L2, March 2017.

\bibitem{fujibayashi2020}
Sho Fujibayashi, Masaru Shibata, Shinya Wanajo, Kenta Kiuchi, Koutarou Kyutoku,
  and Yuichiro Sekiguchi.
\newblock Mass ejection from disks surrounding a low-mass black hole: Viscous
  neutrino-radiation hydrodynamics simulation in full general relativity.
\newblock {\em Physical Review D}, 101(8):083029, 2020.

\bibitem{radice2020}
David Radice.
\newblock Binary neutron star merger simulations with a calibrated turbulence
  model.
\newblock {\em arXiv preprint arXiv:2005.09002}, 2020.

\bibitem{leonard75}
A.~{Leonard}.
\newblock {Energy Cascade in Large-Eddy Simulations of Turbulent Fluid Flows}.
\newblock {\em Advances in Geophysics}, 18:237--248, 1975.

\bibitem{muller02a}
W.-C. {M{\"u}ller} and D.~{Carati}.
\newblock {Dynamic gradient-diffusion subgrid models for incompressible
  magnetohydrodynamic turbulence}.
\newblock {\em Physics of Plasmas}, 9:824--834, March 2002.

\bibitem{vigano19b}
Daniele {Vigan{\`o}}, Ricard {Aguilera-Miret}, and Carlos {Palenzuela}.
\newblock {Extension of the subgrid-scale gradient model for compressible
  magnetohydrodynamics turbulent instabilities}.
\newblock {\em Physics of Fluids}, 31(10):105102, Oct 2019.

\bibitem{carrasco19}
Federico Carrasco, Daniele Vigan{\`o}, and Carlos Palenzuela.
\newblock Gradient subgrid-scale model for relativistic mhd large-eddy
  simulations.
\newblock {\em Physical Review D}, 101(6):063003, 2020.

\bibitem{vigano20}
Daniele {Vigan{\`o}}, Ricard {Aguilera-Miret}, Federico {Carrasco}, Borja
  {Mi{\~n}ano}, and Carlos {Palenzuela}.
\newblock {General relativistic MHD large eddy simulations with gradient
  subgrid-scale model}.
\newblock {\em \prd}, 101(12):123019, June 2020.

\bibitem{bonabook}
C.~{Bona}, C.~{Palenzuela-Luque}, and C.~{Bona-Casas}, editors.
\newblock {\em {Elements of Numerical Relativity and Relativistic
  Hydrodynamics}}, volume 783 of {\em Lecture Notes in Physics, Berlin Springer
  Verlag}, 2009.

\bibitem{alic12}
Daniela {Alic}, Carles {Bona-Casas}, Carles {Bona}, Luciano {Rezzolla}, and
  Carlos {Palenzuela}.
\newblock {Conformal and covariant formulation of the Z4 system with
  constraint-violation damping}.
\newblock {\em \prd}, 85(6):064040, Mar 2012.

\bibitem{bezares17}
Miguel Bezares, Carlos Palenzuela, and Carles Bona.
\newblock Final fate of compact boson star mergers.
\newblock {\em Phys. Rev. D}, 95:124005, Jun 2017.

\bibitem{palenzuela18}
C.~{Palenzuela}, B.~{Mi{\~n}ano}, D.~{Vigan{\`o}}, A.~{Arbona},
  C.~{Bona-Casas}, A.~{Rigo}, M.~{Bezares}, C.~{Bona}, and J.~{Mass{\'o}}.
\newblock {A Simflowny-based finite-difference code for high-performance
  computing in numerical relativity}.
\newblock {\em Classical and Quantum Gravity}, 35(18):185007, September 2018.

\bibitem{arbona13}
A.~{Arbona}, A.~{Artigues}, C.~{Bona-Casas}, J.~{Mass{\'o}}, B.~{Mi{\~n}ano},
  A.~{Rigo}, M.~{Trias}, and C.~{Bona}.
\newblock {Simflowny: A general-purpose platform for the management of physical
  models and simulation problems}.
\newblock {\em Computer Physics Communications}, 184:2321--2331, October 2013.

\bibitem{arbona18}
A.~{Arbona}, B.~{Mi{\~n}ano}, A.~{Rigo}, C.~{Bona}, C.~{Palenzuela},
  A.~{Artigues}, C.~{Bona-Casas}, and J.~{Mass{\'o}}.
\newblock {Simflowny 2: An upgraded platform for scientific modelling and
  simulation}.
\newblock {\em Computer Physics Communications}, 229:170--181, August 2018.

\bibitem{hornung02}
Richard~D. Hornung and Scott~R. Kohn.
\newblock Managing application complexity in the samrai object-oriented
  framework.
\newblock {\em Concurrency and Computation: Practice and Experience},
  14(5):347--368, 2002.

\bibitem{gunney16}
Brian~T.N. Gunney and Robert~W. Anderson.
\newblock Advances in patch-based adaptive mesh refinement scalability.
\newblock {\em Journal of Parallel and Distributed Computing}, 89:65 -- 84,
  2016.

\bibitem{vigano19}
Daniele {Vigan{\`o}}, David {Mart{\'\i}nez-G{\'o}mez}, Jos{\'e}~A. {Pons},
  Carlos {Palenzuela}, Federico {Carrasco}, Borja {Mi{\~n}ano}, Antoni
  {Arbona}, Carles {Bona}, and Joan {Mass{\'o}}.
\newblock {A Simflowny-based high-performance 3D code for the generalized
  induction equation}.
\newblock {\em Computer Physics Communications}, 237:168--183, Apr 2019.

\bibitem{liebling20}
S.~L. {Liebling}, C.~{Palenzuela}, and L.~{Lehner}.
\newblock Toward fidelity and scalability in non-vacuum mergers.
\newblock {\em Classical and Quantum Gravity}, 37(13):135006, jun 2020.

\bibitem{shu98}
Chi-Wang Shu.
\newblock {\em Essentially non-oscillatory and weighted essentially
  non-oscillatory schemes for hyperbolic conservation laws}, pages 325--432.
\newblock Springer Berlin Heidelberg, Berlin, Heidelberg, 1998.

\bibitem{suresh97}
A.~Suresh and H.T. Huynh.
\newblock Accurate monotonicity-preserving schemes with runge–kutta time
  stepping.
\newblock {\em Journal of Computational Physics}, 136(1):83 -- 99, 1997.

\bibitem{McCorquodale:2011}
Peter McCorquodale and Phillip Colella.
\newblock A high-order finite-volume method for conservation laws on locally
  refined grids.
\newblock {\em Commun. Appl. Math. Comput. Sci.}, 6(1):1--25, 2011.

\bibitem{Mongwane:2015}
B.~{Mongwane}.
\newblock {Toward a consistent framework for high order mesh refinement schemes
  in numerical relativity}.
\newblock {\em General Relativity and Gravitation}, 47:60, May 2015.

\bibitem{read09}
J.~S. {Read}, B.~D. {Lackey}, B.~J. {Owen}, and J.~L. {Friedman}.
\newblock Constraints on a phenomenologically parametrized neutron-star
  equation of state.
\newblock {\em Physical Review D}, 79(12), Jun 2009.

\bibitem{bauswein10}
A.~Bauswein, H.~Th. Janka, and R.~Oechslin.
\newblock {Testing Approximations of Thermal Effects in Neutron Star Merger
  Simulations}.
\newblock {\em Phys. Rev.}, D82:084043, 2010.

\bibitem{kastaun20}
Wolfgang {Kastaun}, Jay {Vijay Kalinani}, and Riccardo {Ciolfi}.
\newblock {Robust Recovery of Primitive Variables in Relativistic Ideal
  Magnetohydrodynamics}.
\newblock {\em arXiv e-prints}, page arXiv:2005.01821, May 2020.

\bibitem{lorene}
{\sc Lorene} home page.
\newblock \url{http://www.lorene.obspm.fr/}, 2010.

\bibitem{kazantsev68}
A.~P. {Kazantsev}.
\newblock {Enhancement of a Magnetic Field by a Conducting Fluid}.
\newblock {\em Soviet Journal of Experimental and Theoretical Physics},
  26:1031, May 1968.

\bibitem{miura82}
A.~{Miura} and P.~L. {Pritchett}.
\newblock {Nonlocal stability analysis of the MHD Kelvin-Helmholtz instability
  in a compressible plasma}.
\newblock {\em JGR}, 87:7431--7444, September 1982.

\bibitem{paschalidis2015one}
Vasileios Paschalidis, William~E East, Frans Pretorius, and Stuart~L Shapiro.
\newblock One-arm spiral instability in hypermassive neutron stars formed by
  dynamical-capture binary neutron star mergers.
\newblock {\em Physical Review D}, 92(12):121502, 2015.

\bibitem{east2016relativistic}
William~E East, Vasileios Paschalidis, Frans Pretorius, and Stuart~L Shapiro.
\newblock Relativistic simulations of eccentric binary neutron star mergers:
  One-arm spiral instability and effects of neutron star spin.
\newblock {\em Physical Review D}, 93(2):024011, 2016.

\bibitem{giacomazzo11}
B.~{Giacomazzo}, L.~{Rezzolla}, and L.~{Baiotti}.
\newblock {Accurate evolutions of inspiralling and magnetized neutron stars:
  Equal-mass binaries}.
\newblock {\em \prd}, 83(4):044014, February 2011.

\bibitem{most19}
E.~R. {Most}, L.~Jens {Papenfort}, and L.~{Rezzolla}.
\newblock {Beyond second-order convergence in simulations of magnetized binary
  neutron stars with realistic microphysics}.
\newblock {\em \mnras}, 490(3):3588--3600, Dec 2019.

\bibitem{durran17}
D.~{Durran}, J.~A. {Weyn}, and M.~Q. {Menchaca}.
\newblock {Practical Considerations for Computing Dimensional Spectra from
  Gridded Data}.
\newblock {\em Monthly Weather Review}, 145:3901--3910, September 2017.

\end{thebibliography}

\end{document}